\documentclass[review]{elsarticle}
\makeatletter
\def\ps@pprintTitle{%
 \let\@oddhead\@empty
 \let\@evenhead\@empty
 \def\@oddfoot{\centerline{\thepage}}%
 \let\@evenfoot\@oddfoot}
\makeatother

\usepackage{amssymb}

\usepackage{amsmath}

\usepackage{graphicx}

\usepackage{multirow}
\usepackage{ctable}

\usepackage{placeins}

\usepackage{ulem}

\usepackage{appendix}

\usepackage{setspace}

\usepackage{hyperref}
\usepackage{url}

\begin{document}

\begin{frontmatter}
\title{GPU implementation of algorithm SIMPLE-TS for calculation of unsteady, viscous, compressible and heat-conductive gas flows}
\author{Kiril S. Shterev}
\address{Institute of Mechanics, Bulgarian Academy of Sciences,\\
 Acad. G. Bonchev Str., Block 4, Sofia 1113, Bulgaria}
\ead{kshterev@imbm.bas.bg}
\ead[url]{http://www.imbm.bas.bg}

\begin{abstract}
The recent trend of using Graphics Processing Units (GPU's) for high performance computations is driven by the high ratio of price performance for these units, complemented by their cost effectiveness. At first glance, computational fluid dynamics (CFD) solvers match perfectly to GPU resources because these solvers make intensive calculations and use relatively little memory. Nevertheless, there are scarce results about the practical use of this serious advantage of GPU over CPU, especially for calculations of viscous, compressible, heat-conductive gas flows with double precision accuracy. In this paper, two GPU algorithms according to time approximation of convective terms were presented: explicit and implicit scheme. To decrease data transfers between device memories and increase the arithmetic intensity of a GPU code we minimize the number of kernels. The GPU algorithm was implemented in one kernel for the implicit scheme and two kernels for the explicit scheme. The numerical equations were put together using macros and optimization, data copy from global to private memory, and data reuse were left to the compiler. Thus keeps the code simpler with excellent maintenance. As a test case, we model the flow past squares in a microchannel at supersonic speed. The tests show that overall speedup of AMD Radeon R9 280X is up to 102x compared to Intel Core i5-4690 core and up to 184x compared to Intel Core i7-920 core, while speedup of NVIDIA Tesla M2090 is up to 11x compared to Intel Core i5-4690 core and up to 20x compared to Intel Core i7-920 core. Memory requirements of GPU code are improved compared to CPU one. It requires 1[GB] global memory for 5.9 million finite volumes that are two times less compared to C++ CPU code. After all the code is simple, portable (written in OpenCL), memory efficient and easily modifiable moreover demonstrates excellent performance.
\end{abstract}


\end{frontmatter}

\section{Introduction}
Computational analysis of fluid dynamics problems depends strongly on the computational resources \cite{Versteeg2007}. The computational demands are related mainly to the floating point performance and the memory size.\\\indent
In the last few years, the performance of Graphics Processing Units (GPU's) overcame significantly the performance of Central Processor Units (CPUs), see \cite{Nickolls_2010}, \cite{Brodtkorb2013}. At first glance, computational fluid dynamics (CFD) solvers match perfectly to GPU resources, because these solvers make intensive calculations and use relatively little memory. Nevertheless, there are scarce results about the practical use of this serious advantage of GPU over CPU, especially for calculations of viscous, compressible, heat-conductive gas flows with double precision accuracy. The reported speedups of GPU code to CPU code strongly depend on the mathematical model and the precision of floating point operations. The calculation of Euler flow with single precision in \cite{Elsen2008} demonstrates speedup of over 40x when comparing GPU NVIDIA 8800GTX and CPU Intel Core 2 Duo. The calculation of incompressible fluid demonstrates speedup up to 48 times compared to serial CPU code. Zaspel and Griebel \cite{Zaspel2012} report 4.0x speedup when comparing GPU NVIDIA C2050 and two Intel Xeon X5650 (six cores CPU) that is equivalent to a speedup of 48x when the GPU code is compared to one CPU core (serial code). Cohen and Molemaker \cite{Cohen2009fast} report 8.5x speedup when comparing GPU code run on NVIDIA Quadro FX5800 and CPU parallel code run on an 8-core dual socket Xeon E5420 at 2.5GHz. The equivalent speedup is 42x when the GPU code is compared to one CPU core (serial code). The calculation of compressible fluid reached slower speedup on GPU than Euler and incompressible flow calculations. Salvadore, Bernardini, and Botti demonstrate speedup of 11x when comparing GPU NVIDIA Tesla S2070 and serial code executed on Intel Xeon X5650, \cite{Salvadore2013}. Liang, Liu, and Yuan calculate the seven-equation model for compressible two-phase flow on NVIDIA Tesla C2075 GPU. The GPU code is 31x faster compared with one Intel Xeon Westmere 5675 CPU core, see \cite{Liang2014}. The reported speedups depend on the calculated problem and used hardware.\\\indent
In this paper is presented GPU algorithm for calculation of unsteady, viscous, compressible and heat-conductive gas. The algorithm is a mix of a couple of ideas. The first idea is the minimization of data transfers between memories. We copy all simulation data to the GPU once at the beginning of the application. Therefore, almost no GPU $\leftrightarrow$ CPU data transfers are necessary during the simulation, similar as \cite{Zaspel2012}. Data transfers between global and local device memories are another possible bottleneck. GPU version of algorithm SIMPLE-TS is developed so that minimize number of kernels to one (see Fig. \ref{SIMPLE-TS_for_CPU_and_GPU_implicit}) or two (see see Fig. \ref{SIMPLE-TS_for_CPU_and_GPU_explicit}). The algorithm SIMPLE-TS is developed to be easily parallel organized that makes a possible realization of this minimal kernel concept up to one or two kernels. As a result data transfers between memories of host $\leftrightarrow$ device and global $\leftrightarrow$ local/private memories of the device are minimized. The other idea is to left optimization to the compiler. We put together numerical equations using macros. As a result floating point operations per equation reached up to 388. The optimization, data copy from global to private memory, and data reuse were left to the compiler. Thus keeps the code simpler with excellent maintenance.\\\indent
The proposed concept was applied to different approximation schemes of convective terms. According to time, explicit scheme (Forward Euler) and implicit scheme (Backward Euler) convective terms approximation. On the other side, upwind first order scheme and Total Variation Diminishing (TVD) second order scheme, with Van Leer limiter \cite{vanLeer1974} approximate convective terms by space.\\\indent
The portability of the code is important to run the code on different devices with none or minimal corrections. To this aim GPU code was written in OpenCL (Open Computing Language). OpenCL is a royalty-free standard for cross-platform, parallel programming of modern processors found in personal computers, servers, and handheld/embedded devices (see \cite{OpenCL_standard}). OpenCL implementers are Intel, Texas Instruments, Marvell, Apple Inc., NVIDIA Corporation, MediaTek Inc, QUALCOMM, AMD, Altera Corporation, Vivante Corporation, Xilinx Inc., ARM Limited, Imagination Technologies, STMicroelectronics International NV, IBM Corporation, Creative Labs and Samsung Electronics. One can view a complete list of companies and their conformant products in \cite{OpenCL_companies_and_their_conformant_products}. OpenCL gives the possibility of Portable Heterogeneous programming of diverse compute resources. One code tree can be executed on CPUs, GPU's, DSPs, FPGA, and hardware. Can be organized dynamically interrogate system load and balance work across available processors. One can find out more information about OpenCL programming in \cite{OpenCL_standard}, \cite{OpenCL_Programming_Guide_2011}, \cite{Heterogeneous_Computing_with_OpenCL_2012}. On the other hand, CUDA (Compute Unified Device Architecture) language is widely used from the scientific community to calculate computationally expensive problems, see \cite{CUDA_zone}, \cite{ProgrammingMassivelyParallelProcessors_2010},\cite{Cuda_by_Examples_2010}. Contrary OpenCL the CUDA is supported only by NVIDIA devices, see \cite{CUDA_GPUs}. The GPU code in presented paper was written in OpenCL and performance was obtained on AMD GPU (AMD Radeon R9 280X) and NVIDIA GPU (NVIDIA Tesla M2090). The terminology related to GPU used here is according OpenCL.\\\indent
CPU serial code performance was obtained on CPU Intel Core i7-920 and CPU Intel Core i5-4690 while GPU code performance was obtained on GPU AMD Radeon R9 280X and GPU NVIDIA Tesla M2090. Both codes use double precision floating point operations.
\section{Continuum model equations}\indent
A two dimensional system of equations describing the unsteady flow of viscous, compressible, heat-conductive fluid can be expressed in a general form as follows:
\begin{equation}
\frac{\partial \rho }{\partial t} + \frac{\partial (\rho
u)}{\partial x} + \frac{\partial (\rho v)}{\partial y} = 0
    \label{pl4}
\end{equation}
\begin{equation}
\begin{split}
\frac{\partial(\rho u)}{\partial t} +& \frac{\partial(\rho u u)}{\partial x} + \frac{\partial(\rho v u)}{\partial y}
= \rho g_x - A \frac{\partial p}{\partial x} + B\left[\frac{\partial}{\partial x}\left(\Gamma \frac{\partial u}{\partial x}\right) + \frac{\partial}{\partial y}\left(\Gamma \frac{\partial u}{\partial y}\right)\right] \\
 & + B \left\{\frac{\partial}{\partial x}\left(\Gamma \frac{\partial u}{\partial x}\right) + \frac{\partial}{\partial y}\left(\Gamma \frac{\partial v}{\partial x}\right) - \frac{2}{3} \frac{\partial}{\partial x}\left[\Gamma\left(\frac{\partial u}{\partial x} + \frac{\partial v}{\partial y}\right)\right]\right\}
    \label{pl2}
\end{split}
\end{equation}
\begin{equation}
\begin{split}
\frac{\partial(\rho v)}{\partial t} +& \frac{\partial(\rho u v)}{\partial x} + \frac{\partial(\rho v v)}{\partial y}
= \rho g_y - A \frac{\partial p}{\partial y} + B\left[\frac{\partial}{\partial x}\left(\Gamma \frac{\partial v}{\partial x}\right) + \frac{\partial}{\partial y}\left(\Gamma \frac{\partial v}{\partial y}\right)\right] \\
 & + B \left\{\frac{\partial}{\partial y}\left(\Gamma \frac{\partial v}{\partial y}\right) + \frac{\partial}{\partial x}\left(\Gamma \frac{\partial u}{\partial y}\right) - \frac{2}{3} \frac{\partial}{\partial y}\left[\Gamma\left(\frac{\partial u}{\partial x} + \frac{\partial v}{\partial y}\right)\right]\right\}
    \label{pl3}
\end{split}
\end{equation}
\begin{equation}
\begin{split}
\frac{\partial(\rho T)}{\partial t} +& \frac{\partial(\rho u T)}{\partial x} + \frac{\partial(\rho v T)}{\partial y} \\
=& C^{T1}\left[\frac{\partial}{\partial x}\left(\Gamma^{\lambda}\frac{\partial T}{\partial x}\right) + \frac{\partial}{\partial y}\left(\Gamma^{\lambda}\frac{\partial T}{\partial y}\right)\right]
+ C^{T2}.\Gamma.\Phi + C^{T3}\frac{Dp}{Dt}
    \label{pl6}
\end{split}
\end{equation}
\begin{equation}
    p=\rho T
    \label{pl5}
\end{equation}
where:
\begin{equation}
    \Phi=2\left[\left(\frac{\partial u}{\partial x}\right)^2 + \left(\frac{\partial v}{\partial y}\right)^2\right] + \left(\frac{\partial v}{\partial x} + \frac{\partial u}{\partial y}\right)^2 - \frac{2}{3}\left(\frac{\partial u}{\partial x} + \frac{\partial v}{\partial y}\right)^2
    \label{pl7}
\end{equation}
$u$ is the horizontal component of velocity, $v$ is the vertical component of velocity, $p$ is pressure, $T$ is temperature, $\rho$ is density, $t$ is time, $x$ and  $y$ are coordinates of a Cartesian coordinate system. The parameters $A$, $B$, $g_x$, $g_y$, $C^{T1}$, $C^{T2}$, $C^{T3}$ and diffusion coefficients $\Gamma$ and $\Gamma^\lambda$, given in Eqs. (\ref{pl4})-(\ref{pl5}), depend on the gas model and the equation in non-dimensional form. Upwind first order scheme and Total Variation Diminishing (TVD) second order scheme, with Van Leer limiter, approximate convective terms. A second order central difference scheme approximate diffusion terms. According time explicit scheme (Forward Euler) and implicit scheme (Backward Euler) approximate convective terms, while implicit scheme (Backward Euler) approximate diffusion terms. This approach is a typical approach for approximation of partial differential equations of convective-diffusion type \cite{Ascher1997_implicit-explicit_methods}.
\section{Porting algorithm SIMPLE-TS to GPU}
GPU algorithm development requires an understanding of specifics of the algorithm for implementation and target device, GPU in this case. Firstly is presented GPU implementation of algorithm SIMPLE-TS, subsection \ref{GPU_algorithm_section}. Next subsection \ref{GPU_numerical_equations_section} present implementation of numerical equations. After that is presented GPU specifics that was taken into account, subsections \ref{GPU_specifics_section}. Finally, in this section are presented important tips and tricks that can increase performance significantly, subsection \ref{GPU_tips_section}.
\subsection{Algorithm SIMPLE-TS}\label{GPU_algorithm_section}
The algorithm SIMPLE-TS \cite{Shterev2010} is developed with the idea of easy parallel implementation. It is an iterative Jacobi method; however SIMPLE-TS is faster than SIMPLE \cite{Patankar1972} and PISO \cite{Issa1986} and do not need under-relaxation coefficients to ensure convergence.\\\indent
In the early stage of development of GPU algorithm SIMPLE-TS used only implicit scheme to approximate convective and diffusion terms. The internal loop for calculation of time step was in a single kernel, see loop 2 on Fig. \ref{SIMPLE-TS_for_CPU_and_GPU_implicit}. This work is reported in \cite{Shterev_GPU_2013}. The presented here version of the corresponding algorithm is with improved performance of a couple of times.\\\indent
In this paper are presented and tested algorithm with different approximations of convective terms. According time explicit scheme (Forward Euler) and implicit scheme (Backward Euler) approximate convective terms. On the other side according space upwind first order scheme and Total Variation Diminishing (TVD) second order scheme, with Van Leer limiter approximate convective terms. After all four variants of SIMPLE-TS algorithms are tested, which are noted as follow:
\begin{itemize}
\item explicit TVD second-order scheme - approximate convective terms with explicit (Forward Euler) and TVD second-order scheme
\item explicit upwind first-order scheme - approximate convective terms with explicit (Forward Euler) and upwind first-order scheme
\item implicit TVD second-order scheme - approximate convective terms with explicit (Backward Euler) and TVD second-order scheme
\item implicit upwind first-order scheme - approximate convective terms with explicit (Backward Euler) and upwind first-order scheme
\end{itemize}
Explicit and implicit schemes possess well-known advantages and disadvantages. Explicit scheme compared to the implicit scheme are less stable for fast flows, but are simpler to program and requires less computational efforts. Explicit scheme corresponds very well to TVD second order schemes. TVD schemes are applicable for calculation of steady, slow or moderate fluid flows. They reduce the number of nodes in computation domain significantly, because of it is second-order accuracy in space. On the other side TVD scheme increase the number of floating point operations, see Table \ref{N_FLOP_GPU_explicit} and Table \ref{N_FLOP_GPU_implicit}. Explicit TVD second-order scheme reduces approximately two times floating point operations compared to implicit one. An explicit TVD second-order scheme is recommended for calculation of steady, slow or moderate fluid flows. The fast flows, where explicit TVD second-order scheme obtains physical unrealistic osculations, can be calculated with an implicit upwind first-order scheme. The implicit upwind first-order scheme is the most stable of discussed schemes.
\begin{table}[htb!]
\begin{center}
\begin{tabular}{ p{5cm} p{4cm} p{1.5cm}}
\toprule
  \textbf{Approximation scheme}
& \textbf{Convective terms}
& \textbf{loop 2} \\
\toprule
TVD second-order & 687 & 618 \\
\midrule
upwind first-order & 143 & 486 \\
\bottomrule
\end{tabular}
\caption{Number of floating point operations for GPU algorithm SIMPLE-TS, when convective terms are approximated with explicit scheme and without implementation of boundary conditions in numerical equations.}
\label{N_FLOP_GPU_explicit}
\end{center}
\end{table}
\begin{table}[htb!]
\begin{center}
\begin{tabular}{ p{5cm} p{1.5cm}}
\toprule
  \textbf{Approximation scheme}
& \textbf{loop 2} \\
\toprule
TVD second-order & 1229 \\
\midrule
upwind first-order & 687 \\
\bottomrule
\end{tabular}
\caption{Number of floating point operations for GPU algorithm SIMPLE-TS, when convective terms are approximated with implicit scheme and without implementation of boundary conditions in numerical equations.}
\label{N_FLOP_GPU_implicit}
\end{center}
\end{table}\\\indent
The number of kernels (functions executed on GPU) of explicit and implicit schemes is different. The implicit scheme execute one kernel on GPU that do all calculations of loop 2, see Fig. \ref{SIMPLE-TS_for_CPU_and_GPU_implicit}. The explicit scheme calculate convective terms once per time step. Therefore explicit scheme needs two kernels: one to calculate convective terms and other to calculate loop 2, see Fig. \ref{SIMPLE-TS_for_CPU_and_GPU_explicit}. The convective terms are calculated before loop 2, the results are stored in GPU global memory and reused in loop 2.\\\indent
\begin{figure}[htb!]
    \centering
    \footnotesize

    \begin{minipage}{0.49\textwidth}
        \centering
        \setstretch{1}
		\textbf{CPU (serial)}\\
        \raggedright
            \textbf{Initialize variables.}\\
            \textbf{Start loop 1}:
            \begin{description}
                \item Set the initial condition for the calculated time step.
                \item
                \item \ \\[-1.2mm]
                \item Calculate convective terms of velocities and temperature equations, which are approximated with explicit scheme.
                \item \ \\[5.4mm]
                \item \textbf{Start loop 2} (calculate a state for a new time step):
                \begin{description}
                    \item Calculate diffusion fluxes.
                    \item Calculate pseudo velocities (velocities, without pressure term), coefficients for pressure equation.
                    \item \textbf{Start loop 3}:
                    \begin{description}
                        \item Calculate the coupled equations for energy and pressure.
                    \end{description}
                    \item \textbf{Stop loop 3}. In most cases two iterations are sufficient.
                    \item Calculate velocities using pseudo velocities and pressure (calculated within loop 3).
                \item Compute density, using pressure and temperature calculated within loop 3.
                \item \textbf{Convergence of loop 2}: Check for convergence of the iteration process for the current time step.
                \end{description}
                \item \textbf{Convergence of loop 1}: If the final time is not reached continue.
            \end{description}
    \end{minipage}
    \begin{minipage}{0.49\textwidth}
        \centering
        \setstretch{1}
        \textbf{GPU code}\\
        \raggedright
            \textbf{Initialize variables.}\\
            \textbf{Start loop 1}:
            \begin{description}
                \item Set the initial condition for the calculated time step.
                \item \textbf{Queue kernel for execution on GPU to calculate convective terms:}
                    \item Calculate convective terms of velocities and temperature equations, approximated with explicit scheme and store data in global device memory.
                \item \textbf{Finish kernel execution.}
                \item \textbf{Start loop 2} (calculating a state for a new time step):
                \begin{description}
                    \item \textbf{Queue kernel for execution on GPU to do calculations in loop 2:}
                        \item Calculate equation for energy.
                        \item Calculate pseudo velocities (velocities, without pressure term), coefficients for pressure equation and store data in local memory.
                        \item Calculate equation for pressure.
                        \item
                        \item Calculate velocities using pseudo velocities and pressure.
                    \item \textbf{Finish kernel execution.}
                    \item
                    \item
                \item \textbf{Convergence of loop 2}: Check for convergence of the iteration process for the current time step.
                \end{description}
                \item \textbf{Convergence of loop 1}: If the final time is not reached continue.
            \end{description}
    \end{minipage}
    \caption{Algorithm SIMPLE-TS for CPU (serial) and GPU that use explicit approximation scheme for convective terms.}
    \label{SIMPLE-TS_for_CPU_and_GPU_explicit}
    \normalsize
\end{figure}

\begin{figure}[htb!]
    \centering
    \footnotesize
    \begin{minipage}{0.49\textwidth}
        \centering
        \textbf{CPU (serial) \cite{Shterev2010}}\\
        \raggedright
            \textbf{Initialize variables.}\\
            \textbf{Start loop 1}:
            \begin{description}
                \item Set the initial condition for the calculated time step.
                \item \textbf{Start loop 2} (calculating a state for a new time step):
                \begin{description}
                    \item Calculate convective and diffusion fluxes.
                    \item Calculate pseudo velocities (velocities, without pressure term), coefficients for pressure equation.
                    \item \textbf{Start loop 3}:
                    \begin{description}
                        \item Calculate the coupled equations for energy and pressure.
                    \end{description}
                    \item \textbf{Stop loop 3}. In most cases two iterations are sufficient.
                    \item Calculate velocities using pseudo velocities and pressure (calculated within loop 3).
                \item Compute density, using pressure and temperature calculated within loop 3.
                \item \textbf{Convergence of loop 2}: Check for convergence of the iteration process for the current time step.
                \end{description}
                \item \textbf{Convergence of loop 1}: If the final time is not reached continue.
            \end{description}
    \end{minipage}
    \begin{minipage}{0.49\textwidth}
        \centering
        \textbf{GPU code}\\
        \raggedright
            \textbf{Initialize variables.}\\
            \textbf{Start loop 1}:
            \begin{description}
                \item Set the initial condition for the calculated time step.
                \item \textbf{Start loop 2} (calculating a state for a new time step):
                \begin{description}
                    \item \textbf{Run a kernel on GPU:}
                    \begin{description}
                        \item Calculate equation for energy.
                        \item Calculate pseudo velocities (velocities, without pressure term), coefficients for pressure equation and store data in local memory.
                        \item Calculate equation for pressure.
                        \item
                        \item
                        \item Calculate velocities using pseudo velocities and pressure.
                    \end{description}
                    \item \textbf{Finish kernel execution.}
                    \item
                    \item \ \\[0mm]
                \item \textbf{Convergence of loop 2}: Check for convergence of the iteration process for the current time step.
                \end{description}
                \item \textbf{Convergence of loop 1}: If the final time is not reached continue.
            \end{description}
    \end{minipage}
    \caption{Algorithm SIMPLE-TS for CPU (serial) and GPU that use implicit approximation scheme for convective terms.}
    \label{SIMPLE-TS_for_CPU_and_GPU_implicit}
\end{figure}
\begin{figure}[htb!]
    \centering
    \includegraphics[keepaspectratio=true, width=0.7\textwidth]{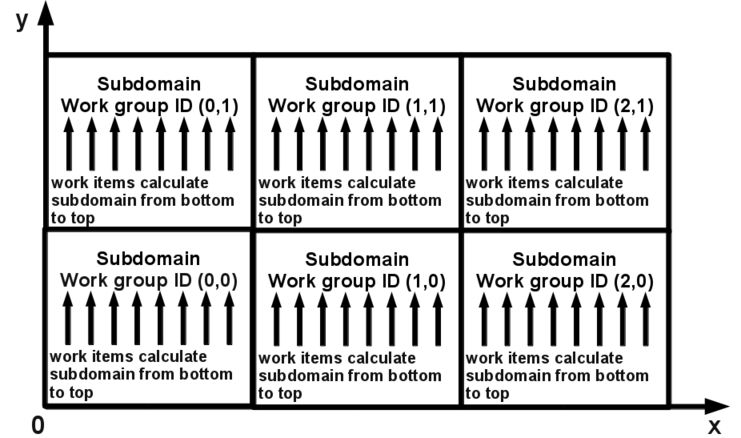}
    \caption{Domain decomposition of computational domain.}
    \label{GPU_Work_group_organization}
\end{figure}

\FloatBarrier
A domain decomposition (data partitioning) approach is used to separate calculations between work groups. A subdomain corresponds to each work group, Fig. \ref{GPU_Work_group_organization}. Brandvik and Pullan \cite{Brandvik_and_Pullan_2008} calculate each subdomain by fix index of x-axes and z-axes to a work item (thread) of a group and doing iteration in kernel along y-axes. They keep neighbors data in local memory (shared memory, according CUDA terminology) and write results straight to the global memory. This approach is partially adopted in presented algorithm. The difference is that we keep temporary arrays in local memory, while neighbors data was copied straight from global memory. Iteration in kernel along the y-axis have important advantages:
\begin{itemize}
\item keep temporary variables as pseudo velocities and pressure coefficients in local memory, instead of slower global memory. This decrease global memory usage and global $\leftrightarrow$ local memory data transfers.
\item increase number of floating point operations in a kernel (see Table \ref{N_FLOP_GPU_explicit} and Table \ref{N_FLOP_GPU_implicit}) and reuse copied data from global memory to private memory.
\end{itemize}
On the other hand, some preliminary calculations of a subdomain have to be done. After all the influence of number of rows per sub-domain over performance is important and is investigated in Section \ref{Speedup_analysis}, further in paper.\\\indent
\FloatBarrier
\subsection{Numerical equations}\label{GPU_numerical_equations_section}
\begin{figure}[htb!]
    \centering
    \includegraphics[width=0.9\textwidth]{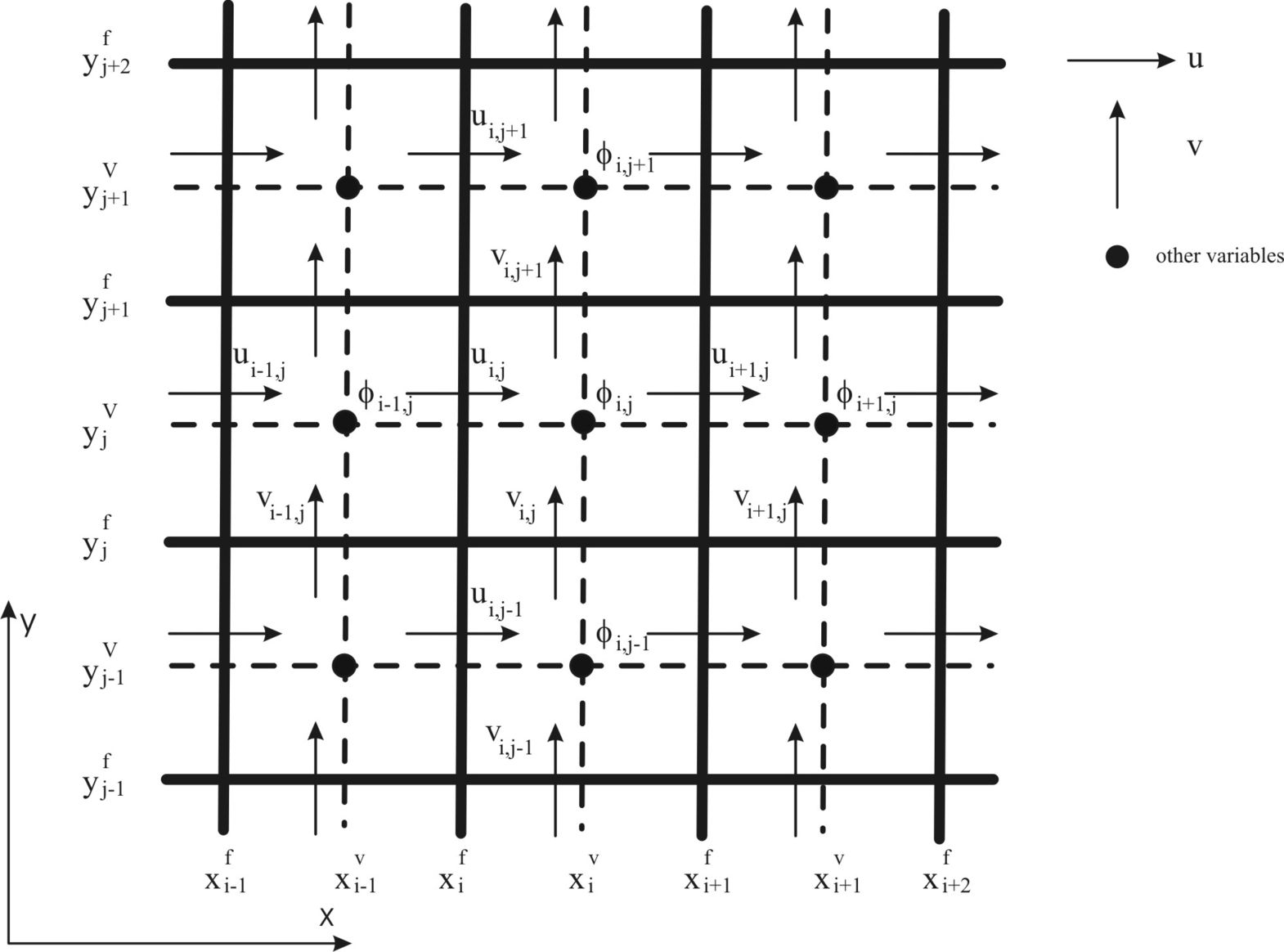}
    \caption{Cell volume}
    \label{CellVolume}
\end{figure}

\begin{figure}[htb!]
    \centering
    \includegraphics[width=1\textwidth]{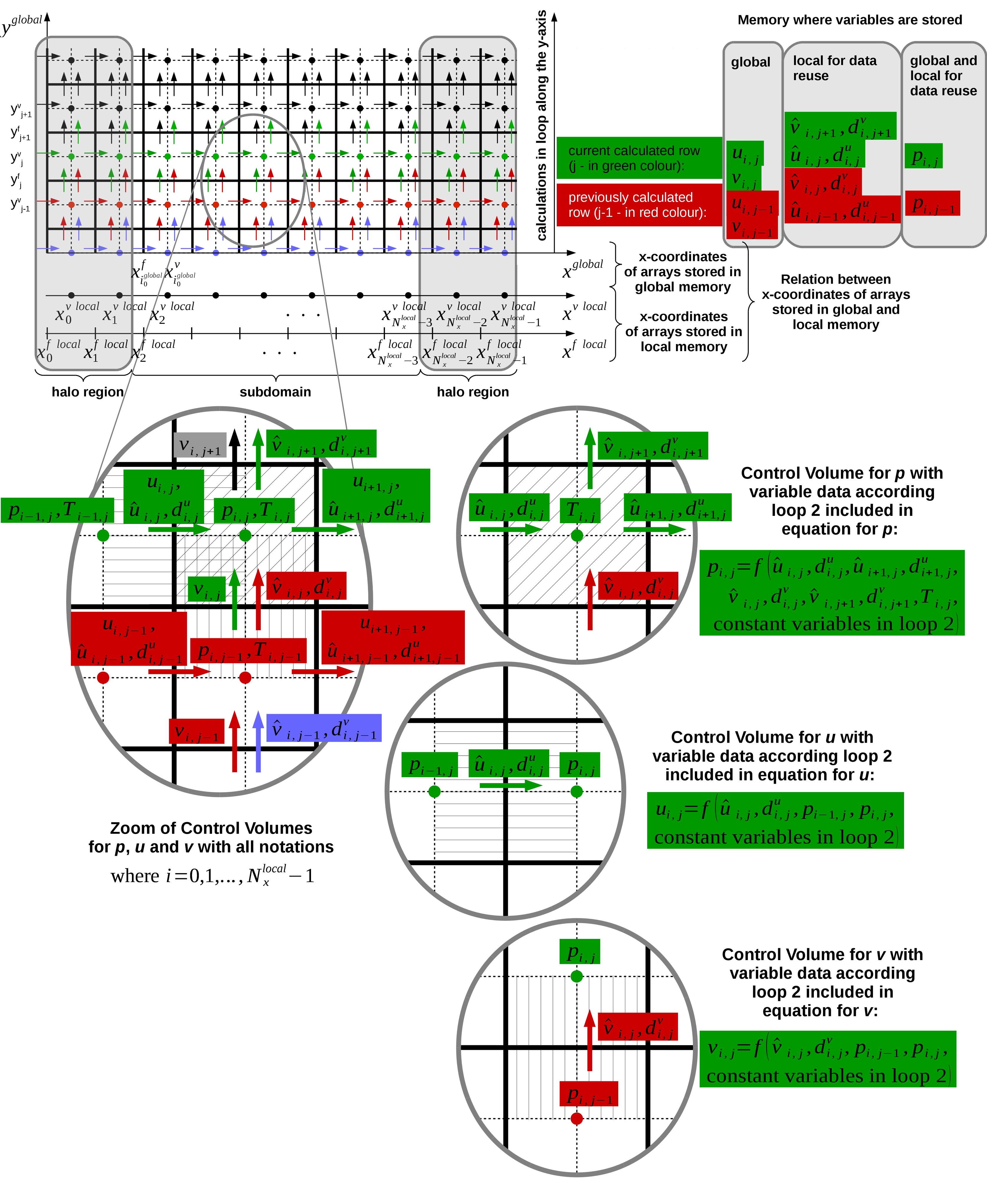}
    \caption{GPU dependent variables $\hat u$, $d^u$, $\hat v$, $d^v$, $p$, $u$, and $v$ calculations.}
    \label{GPU_dependent_variables_calculation}
\end{figure}

Mapping SIMPLE-TS to GPU needs to take into account domain decomposition, sequence of the nodes in mesh calculation and device specifics. SIMPLE-TS and numerical equations derivation are presented in details in \cite{Shterev2010} while here are analysed and mapped to GPU.\\\indent
The algorithm has to correspond to GPU architecture specific to reach good performance. GPU device architecture use Single Instruction, Multiple Data (SIMD) parallelism, in which multiple processors execute the same instructions on different pieces of data. Therefore, the algorithm SIMPLE-TS have to be reorganised as fully Jacobi iterative solver. Therefore, right-hand side of numerical equations has to contain only constant data according to the current iteration of loop 2. The equations for pressure and velocities are coupled and needs a detailed analysis. To this aim are presented the definition of variables in the control volume (Fig. \ref{CellVolume}) and numerical equations used in algorithm SIMPLE-TS (equations from (\ref{pl8}) to (\ref{pl31_1})).\\\indent
\FloatBarrier
In Fig. \ref{CellVolume} the following grid variables are denoted:\\
$x^{f}_{i}$ - the left frontier x-coordinate of the control volume of node $i$ \\
$y^{f}_{j}$ - bottom frontier y-coordinate of the control volume of node $j$\\
$\Delta x_i = x^{f}_{i+1} - x^{f}_{i}$ - step on OX\\
$\Delta y_j = y^{f}_{j+1} - y^{f}_{j}$ - step on OY\\
$x^{v}_{i}$ - x-coordinate of the centre of the control volume of node $i$, $x^{v}_{i} = x^{f}_{i} + 0.5\Delta x_i$\\
$y^{v}_{j}$ - y-coordinate of the centre of the control volume of node $j$, $y^{v}_{j} = y^{f}_{j} + 0.5\Delta y_j$\\
$\phi_{i,j}$ - field variables defined at point $(x^{v}_{i}, y^{v}_{j})$, pressure, temperature, density, diffusion coefficient and etc.\\
$u_{i,j}$ - the horizontal component of velocity defined at point $(x^{f}_{i}, y^{v}_{j})$\\
$v_{i,j}$ - the vertical component of velocity defined at point $(x^{v}_{i}, y^{f}_{j})$\\
$F^{x}_{i,j}$ - the convective mass flux through the surface between
control volumes $(i-1,j)$ and $(i,j)$ (in horizontal direction)
\begin{equation}
    F^{x}_{i,j} = \rho^{u}_{i,j} u^{old}_{i,j} \Delta y_j
    \label{pl8}
\end{equation}
$F^{y}_{i,j}$ - the convective mass flux through the surface between
control volumes $(i,j-1)$ and $(i,j)$ (in vertical direction)
\begin{equation}
    F^{y}_{i,j} = \rho^{v}_{i,j} v^{old}_{i,j} \Delta x_i,
    \label{pl9}
\end{equation}
where:
\begin{equation}
    \rho^{u}_{i,j} =
        \begin{cases}
            \rho^{old}_{i-1,j} + \psi_s \left(\rho^{old}_{i-2,j},\rho^{old}_{i-1,j},\rho^{old}_{i,j},\rho^{old}_{i+1,j},\Delta x_{i-2},\Delta x_{i-1},\Delta x_{i},\Delta x_{i+1},u^{old}_{i,j} \right)& \text{if $u_{i,j} > 0$},\\
            \rho^{old}_{i,j} + \psi_s \left(\rho^{old}_{i-2,j},\rho^{old}_{i-1,j},\rho^{old}_{i,j},\rho^{old}_{i+1,j},\Delta x_{i-2},\Delta x_{i-1},\Delta x_{i},\Delta x_{i+1},u^{old}_{i,j} \right) & \text{otherwise}.
        \end{cases}
    \label{pl10}
\end{equation}
\begin{equation}
    \rho^{v}_{i,j} =
        \begin{cases}
            \rho^{old}_{i,j-1} + \psi_s \left(\rho^{old}_{i,j-2},\rho^{old}_{i,j-1},\rho^{old}_{i,j},\rho^{old}_{i,j+1},\Delta y_{j-2},\Delta y_{j-1},\Delta y_{j},\Delta y_{j+1},v^{old}_{i,j} \right)& \text{if $v_{i,j} > 0$},\\
            \rho^{old}_{i,j} + \psi_s \left(\rho^{old}_{i,j-2},\rho^{old}_{i,j-1},\rho^{old}_{i,j},\rho^{old}_{i,j+1},\Delta y_{j-2},\Delta y_{j-1},\Delta y_{j},\Delta y_{j+1},v^{old}_{i,j} \right)& \text{otherwise}.
        \end{cases}
    \label{pl11}
\end{equation}
Density $(\rho)$ is computed at the middle points, i.e. $\rho^{u}_{i,j}$ at point $(x^{f}_{i}, y^{v}_{j})$ and $\rho^{v}_{i,j}$ at point $(x^{v}_{i}, y^{f}_{j})$, by using first order upwind scheme or second order TVD scheme according to convective terms approximation. The implementation of TVD scheme corresponds to presented in \cite{Versteeg2007}. The functions $\psi^c_{i,j}$ and $\psi^s_{i,j}$ approximate TVD correction to the first order upwind scheme at center or surface of the control volumes, respectively. $\psi^c_{i,j}$ is defined at point $(x^v_i, y^v_j)$ while $\psi^s_{i,j}$ is defined at point $(x^f_i, y^v_j)$:
\begin{equation}
\psi_c(\phi_1, \phi_2, \phi_3, \phi_4, \Delta_1, \Delta_2, \Delta_3, v) =
\begin{cases}
	0.5 \psi\left(\frac{\Delta_2(\phi_2-\phi_1)}{\Delta_1 (\phi_3-\phi_2)}\right)& \text{if $v > 0$},\\
	-0.5 \psi\left(\frac{\Delta_2(\phi_4-\phi_3)}{\Delta_3 (\phi_3-\phi_2)}\right)& \text{otherwise}.
\end{cases}
\label{pl15_1}
\end{equation}
\begin{equation}
\psi_s(\phi_1, \phi_2, \phi_3, \phi_4, \Delta_1, \Delta_2, \Delta_3,\Delta_4, v) =
\begin{cases}
	 \frac{\Delta_2}{\Delta_2+\Delta_3} \psi\left(\frac{(\Delta_2+\Delta_3) (\phi_2-\phi_1)}{(\Delta_1+\Delta_2)(\phi_3-\phi_2)}\right)& \text{if $v > 0$},\\
	-\frac{\Delta_3}{\Delta_2+\Delta_3} \psi\left(\frac{(\Delta_2+\Delta_3) (\phi_4-\phi_3)}{(\Delta_3+\Delta_4) (\phi_3-\phi_2)}\right)& \text{otherwise}.
\end{cases}
\label{pl15_2}
\end{equation}
where $\psi(r)$ is TVD limiter. Here is used Van Leer \cite{vanLeer1974} TVD limiter: $\psi(r) = (r+|r|)/(1+r)$. When convective terms and density in middle points are approximated using upwind first order scheme TVD corrections are null ($\psi_c=0$ and $\psi_s=0$).\\\indent
SIMPLE-TS use pseudo velocities in the same way as SIMPLER. Therefore, the numerical equations for $u$ and $v$ can be written in form:
\begin{equation}
    u_{i,j} = \hat{u}_{i,j} - d^u_{i,j}\left(p_{i,j} - p_{i-1,j}\right)
    \label{pl18}
\end{equation}
\begin{equation}
    v_{i,j} = \hat{v}_{i,j} - d^v_{i,j}\left(p_{i,j} - p_{i,j-1}\right)
    \label{pl19}
\end{equation}
where $\hat{u}_{i,j}$ and $\hat{v}_{i,j}$ are pseudo velocities:
\begin{equation}
    \hat{u}_{i,j} = \frac{a^u_1 u^{old}_{i-1,j} + a^u_2 u^{old}_{i+1,j} + a^u_3 u^{old}_{i,j-1} + a^u_4 u^{old}_{i,j+1} + b^u + u^{explicit}_{i,j}}{a^u_0}
    \label{pl20}
\end{equation}
\begin{equation}
    \hat{v}_{i,j} = \frac{a^v_1 v^{old}_{i-1,j} + a^v_2 v^{old}_{i+1,j} + a^v_3 v^{old}_{i,j-1} + a^v_4 v^{old}_{i,j+1} + b^v + v^{explicit}_{i,j}}{a^v_0}
    \label{pl21}
\end{equation}
Where the terms $u^{explicit}_{i,j}$ and $v^{explicit}_{i,j}$ contain explicit approximation of the convective terms.\\\indent
For brevity, here the coefficients for $v$ are given only:
\begin{equation}
\begin{split}
    b^v =& \left(\frac{p^{n-1}_{i,j}}{T^{n-1}_{i,j}} \Delta y_j + \frac{p^{n-1}_{i,j-1}}{T^{n-1}_{i,j-1}} \Delta y_{j-1}\right)\frac{\Delta x_i}{2 \Delta t}v^{n-1}_{i,j} \\
		&+ B \left(\Gamma \!\!\mid_{x^f_{i+1}} (u^{old}_{i+1,j} - u^{old}_{i+1,j-1}) - \Gamma \!\!\mid_{x^f_{i}} (u^{old}_{i,j} - u^{old}_{i,j-1}) \right. \\
		&\ \ \ \ \ \ \ \left. - \frac{2}{3} \Gamma_{i,j} (u^{old}_{i+1,j} - u^{old}_{i,j})
		+ \frac{2}{3} \Gamma_{i,j-1} (u^{old}_{i+1,j-1} - u^{old}_{i,j-1}) \right)
,\\
    d^v_{i,j} =& \frac{A}{a^v_0}\Delta x_i,\\
	\overline{F}^y_{i,j} =& \overline{v}^{old}_{i,j} \rho^{old}_{i,j} \Delta x_{i},\\
	\overline{v}^{old}_{i,j} =& 0.5 \left(v^{old}_{i,j-1} + v^{old}_{i,j} \right).
    \label{pl14}
\end{split}
\end{equation}
The following coefficients correspond to implicit approximation scheme of convective terms:
\begin{equation}
\begin{split}
    a^v_0 =& a^v_1 + a^v_2 + a^v_3 + a^v_4 + 0.5 \left(F^x_{i+1,j} - F^x_{i,j} + F^x_{i+1,j-1} - F^x_{i,j-1}\right) \\
           & + \overline{F}^y_{i,j} - \overline{F}^y_{i,j-1} + \left(\rho^{old}_{i,j} \Delta y_j + \rho^{old}_{i,j-1} \Delta y_{j-1} \right) \frac{\Delta x_i}{2 \Delta t} \\
    a^v_1 =& a^{vc}_1 + D^{vx}_{i,j},\ a^v_2 = a^{vc}_2 + D^{vx}_{i+1,j},\ a^v_3 = a^{vc}_3 + \frac{4}{3} D^{vy}_{i,j},\ a^v_4 = a^{vc}_4 + \frac{4}{3} D^{vy}_{i,j+1} \\
    a^{vc}_0 =& a^{vc}_1 + a^{vc}_2 + a^{vc}_3 + a^{vc}_4 + 0.5(F^x_{i+1,j} - F^x_{i,j} + F^x_{i+1,j-1} - F^x_{i,j-1}) + \overline{F}^y_{i,j} - \overline{F}^y_{i,j-1} \\
    a^{vc}_1 =& 0.5\left[max\left(0, F^x_{i,j}\right) \right.\\
			  &\left. \ \ \ \ \ - F^x_{i,j} \psi_s(v^{old}_{i-2,j},v^{old}_{i-1,j},v^{old}_{i,j},v^{old}_{i+1,j},\Delta x_{i-2},\Delta x_{i-1},\Delta x_{i},\Delta x_{i+1},u^{old}_{i,j}) \right.\\
			&\left.\ \ \ \ \ +max\left(0,F^x_{i,j-1}\right) \right.\\
			  &\left. \ \ \ \ \ - F^x_{i,j-1} \psi_s(v^{old}_{i-2,j},v^{old}_{i-1,j},v^{old}_{i,j},v^{old}_{i+1,j},\Delta x_{i-2},\Delta x_{i-1},\Delta x_{i},\Delta x_{i+1},u^{old}_{i,j-1}) \right]\\
    a^{vc}_2 =& 0.5\left[max\left(0, -F^x_{i+1,j}\right) \right.\\
			  &\left. \ \ \ \ \ - F^x_{i+1,j} \psi_s(v^{old}_{i-1,j},v^{old}_{i,j},v^{old}_{i+1,j},v^{old}_{i+2,j},\Delta x_{i-1},\Delta x_{i},\Delta x_{i+1},\Delta x_{i+2},u^{old}_{i+1,j}) \right.\\
			&\left.\ \ \ \ \ + max\left(0, -F^x_{i+1,j-1}\right) \right.\\
			  &\left. \ \ \ \ \ - F^x_{i+1,j-1} \psi_s(v^{old}_{i-1,j},v^{old}_{i,j},v^{old}_{i+1,j},v^{old}_{i+2,j},\Delta x_{i-1},\Delta x_{i},\Delta x_{i+1},\Delta x_{i+2},u^{old}_{i+1,j-1})\right] \\
    a^{vc}_3 =& max\left(0, \overline{F}^y_{i,j-1}\right) - \overline{F}^y_{i,j-1} \psi_c(v^{old}_{i,j-2},v^{old}_{i,j-1},v^{old}_{i,j},v^{old}_{i,j+1},\Delta y_{j-2},\Delta y_{j-1},\Delta y_{j},\overline{v}^{old}_{i,j}),\\
    a^{vc}_4 =& max\left(0,-\overline{F}^y_{i,j}\right) - \overline{F}^y_{i,j} \psi_c(v^{old}_{i,j-1},v^{old}_{i,j},v^{old}_{i,j+1},v^{old}_{i,j+2},\Delta y_{j-1},\Delta y_{j},\Delta y_{j+1},\overline{v}^{old}_{i,j+1}),\\
    v^{explicit}_{i,j} =& 0.
    \label{pl15}
\end{split}
\end{equation}
The terms $\overline{F}^y_{i,j}$ and $\overline{v}^{old}_{i,j} = 0.5(v^{old}_{i,j} + v^{old}_{i,j+1})$ are defined at point $(x^v_i, y^v_j)$. With $D$ we denote the diffusion conductance at cell face. To determine the value of $D$, we assume that the diffusion $\Gamma$ varies continuously between the adjacent control volumes and use the bilinear interpolation of the diffusion coefficients at the control volume surfaces to solve $\Gamma^{old} \!\!\mid_{x^f_i}$ in $D^{vx}_{i,j}$. For $D^{vy}_{i,j}$ no interpolation is needed, because the diffusion coefficient $\Gamma^{old}$ is defined in nodes $(x^v_i, y^v_j)$.
\begin{equation}
    D^{vx}_{i,j} = B . \Gamma^{old} \!\!\mid_{x^f_i} \frac{\Delta y_j + \Delta y_{j-1}}{\Delta x_i + \Delta x_{i-1}},\ D^{vy}_{i,j} = B . \Gamma^{old}_{i,j-1} \frac{\Delta x_i}{\Delta y_{j-1}}
    \label{pl16}
\end{equation}
The following coefficients correspond to explicit approximation scheme of convective terms:
\begin{equation}
\begin{split}
    a^v_0 =& a^v_1 + a^v_2 + a^v_3 + a^v_4 + \left(\rho^{old}_{i,j} \Delta y_j + \rho^{old}_{i,j-1} \Delta y_{j-1} \right) \frac{\Delta x_i}{2 \Delta t} \\
    a^v_1 =& D^{vx}_{i,j},\ a^v_2 = D^{vx}_{i+1,j},\ a^v_3 = \frac{4}{3} D^{vy}_{i,j},\ a^v_4 = \frac{4}{3} D^{vy}_{i,j+1}\\
    v^{explicit}_{i,j} =&- 0.5 \left[ F^x_{i+1,j-1} (\text{upwind}(v^{n-1}_{i,j}, v^{n-1}_{i+1,j}, u^{n-1}_{i+1,j-1}) \right.\\
	&\left. \ \ \ \ \ \ \ \ \ \ \ \ + (v^{n-1}_{i+1,j} - v^{n-1}_{i,j}) \psi_s(v^{n-1}_{i-1,j},v^{n-1}_{i,j},v^{n-1}_{i+1,j},v^{n-1}_{i+2,j},\Delta x_{i-1},\Delta x_{i},\Delta x_{i+1},\Delta x_{i+2},u^{n-1}_{i+1,j-1}))\right.\\
	&\ \ \ \ \ \ \ \ \left. + F^x_{i+1,j} (\text{upwind}(v^{n-1}_{i,j}, v^{n-1}_{i+1,j}, u^{n-1}_{i+1,j}) \right.\\
	&\left. \ \ \ \ \ \ \ \ \ \ \ \ + (v^{n-1}_{i+1,j} - v^{n-1}_{i,j}) \psi_s(v^{n-1}_{i-1,j},v^{n-1}_{i,j},v^{n-1}_{i+1,j},v^{n-1}_{i+2,j},\Delta x_{i-1},\Delta x_{i},\Delta x_{i+1},\Delta x_{i+2},u^{n-1}_{i+1,j}))\right]\\
	&+ 0.5 \left[ F^x_{i,j-1} (\text{upwind}(v^{n-1}_{i-1,j}, v^{n-1}_{i,j}, u^{n-1}_{i,j-1}) \right.\\
	&\left. \ \ \ \ \ \ \ \ \ \ \ \ + (v^{n-1}_{i,j} - v^{n-1}_{i-1,j}) \psi_s(v^{n-1}_{i-2,j},v^{n-1}_{i-1,j},v^{n-1}_{i,j},v^{n-1}_{i+1,j},\Delta x_{i-2},\Delta x_{i-1},\Delta x_{i},\Delta x_{i+1},u^{n-1}_{i,j-1}))\right.\\
	&\ \ \ \ \ \ \ \ \left. + F^x_{i,j} (\text{upwind}(v^{n-1}_{i-1,j}, v^{n-1}_{i,j}, u^{n-1}_{i,j}) \right.\\
	&\left. \ \ \ \ \ \ \ \ \ \ \ \ + (v^{n-1}_{i,j} - v^{n-1}_{i-1,j}) \psi_s(v^{n-1}_{i-2,j},v^{n-1}_{i-1,j},v^{n-1}_{i,j},v^{n-1}_{i+1,j},\Delta x_{i-2},\Delta x_{i-1},\Delta x_{i},\Delta x_{i+1},u^{n-1}_{i,j}))\right]\\
	& - \Delta x_{i} \rho^{n-1}_{i,j} \overline{v}^{n-1}_{i,j+1} \left[\text{upwind}(v^{n-1}_{i,j}, v^{n-1}_{i,j+1}, \overline{v}^{n-1}_{i,j+1}) \right.\\
	&\left. \ \ \ \ \ \ \ \ \ \ \ \ \ \ \ \ + (v^{n-1}_{i,j+1} - v^{n-1}_{i,j}) \psi_c(v^{n-1}_{i,j-1},v^{n-1}_{i,j},v^{n-1}_{i,j+1},v^{n-1}_{i,j+2},\Delta y_{j-1},\Delta y_{j}, \Delta y_{j+1},\overline{v}^{n-1}_{i,j+1}) \right]\\
	& + \Delta x_{i} \rho^{n-1}_{i,j-1} \overline{v}^{n-1}_{i,j} \left[\text{upwind}(v^{n-1}_{i,j-1}, v^{n-1}_{i,j}, \overline{v}^{n-1}_{i,j}) \right.\\
	&\left. \ \ \ \ \ \ \ \ \ \ \ \ \ \ \ \ + (v^{n-1}_{i,j} - v^{n-1}_{i,j-1}) \psi_c(v^{n-1}_{i,j-2},v^{n-1}_{i,j-1},v^{n-1}_{i,j},v^{n-1}_{i,j+1},\Delta y_{j-2},\Delta y_{j-1}, \Delta y_{j},\overline{v}^{n-1}_{i,j}) \right],
    \label{pl15_11}
\end{split}
\end{equation}
where:
\begin{equation}
    \text{upwind}(\phi_1,\phi_2,v) =
        \begin{cases}
            \phi_1 & \text{if $v > 0$},\\
            \phi_2 & \text{otherwise}.
        \end{cases}
    \label{pl15_12}
\end{equation}
$F^x$ and $F^y$ calculated in explicit terms use previous time step values. $v^{explicit}_{i,j}$ is calculated in separate kernel together with $u^{explicit}_{i,j}$ and $T^{explicit}_{i,j}$ (\ref{pl31_1}) before loop 2 (see Fig. \ref{SIMPLE-TS_for_CPU_and_GPU_explicit}). The results are stored in global device memory and used as constant variables in loop 2.\\\indent
The numerical equation for pressure is expressed as follows:
\begin{equation}
    a^p_0 p_{i,j} = \left(a^{px}_{i,j}p^{old}_{i-1,j} + a^{px}_{i+1,j}p^{old}_{i+1,j} + a^{py}_{i,j}p^{old}_{i,j-1} + a^{py}_{i,j+1}p^{old}_{i,j+1}\right)\Delta t + b^p,
    \label{pl23}
\end{equation}
where:
\begin{equation}
\begin{split}
    &a^p_0 = \frac{1}{T_{i,j}}\Delta x_i \Delta y_j + \left(a^{px}_{i,j} + a^{px}_{i+1,j} + a^{py}_{i,j} + a^{py}_{i,j+1}\right)\Delta t \\
    &b^p = \frac{p^{n-1}_{i,j}}{T^{n-1}_{i,j}}\Delta x_i \Delta y_j - \left(b^{px}_{i+1,j} - b^{px}_{i,j} + b^{py}_{i,j+1} - b^{py}_{i,j}\right)\Delta t \\
    &a^{px}_{i,j} = \rho^u_{i,j}d^u_{i,j}\Delta y_j,\ a^{py}_{i,j} = \rho^v_{i,j}d^v_{i,j}\Delta x_i \\
    &b^{px}_{i,j} = \rho^u_{i,j}\hat{u}_{i,j}\Delta y_j,\ b^{py}_{i,j} = \rho^v_{i,j}\hat{v}_{i,j}\Delta x_i
    \label{pl24}
\end{split}
\end{equation}
The discrete equation for temperature is:
\begin{equation}
\begin{split}
    a^T_0 T_{i,j} =& \Delta t \left(a^T_1 T^{old}_{i-1,j} + a^T_2 T^{old}_{i+1,j} + a^T_3 T^{old}_{i,j-1} + a^T_4 T^{old}_{i,j+1} + S^T_{c_{i,j}} + T^{explicit}_{i,j} \right) \\
    & + \rho^{n-1}_{i,j} T^{n-1}_{i,j} \Delta x_i \Delta y_j,
    \label{pl30}
\end{split}
\end{equation}
where the coefficients that correspond to implicit approximation scheme of convective terms are:
\begin{equation}
\begin{split}
    a^T_0 =& \Delta t \left(a^T_1 + a^T_2 + a^T_3 + a^T_4 + F^x_{i+1,j} - F^x_{i,j} + F^y_{i,j+1} - F^y_{i,j}\right) \\
           & + \rho^{old}_{i,j} \Delta x_i \Delta y_j \\
    a^T_1 =& max\left(0, F^x_{i,j}\right) - F^x_{i,j} \psi_s \left(T^{old}_{i-2,j},T^{old}_{i-1,j},T^{old}_{i,j},T^{old}_{i+1,j},\Delta x_{i-2},\Delta x_{i-1},\Delta x_{i},\Delta x_{i+1},u^{old}_{i,j}\right)\\
		  &+ D^{Tx}_{i,j},\\
    a^T_2 =& max\left(0, -F^x_{i+1,j}\right) - F^x_{i+1,j} \psi_s \left(T^{old}_{i-1,j},T^{old}_{i,j},T^{old}_{i+1,j},T^{old}_{i+2,j},\Delta x_{i-1},\Delta x_{i},\Delta x_{i+1},\Delta x_{i+2},u^{old}_{i+1,j}\right)\\
		  &+ D^{Tx}_{i+1,j},\\
    a^T_3 =& max\left(0, F^y_{i,j}\right) - F^y_{i,j} \psi_s \left(T^{old}_{i,j-2},T^{old}_{i,j-1},T^{old}_{i,j},T^{old}_{i,j+1},\Delta y_{j-2},\Delta y_{j-1},\Delta y_{j},\Delta y_{j+1},v^{old}_{i,j}\right)\\
		  &+ D^{Ty}_{i,j},\\
    a^T_4 =& max\left(0, -F^y_{i,j+1}\right) - F^y_{i,j+1} \psi_s \left(T^{old}_{i,j-1},T^{old}_{i,j},T^{old}_{i,j+1},T^{old}_{i,j+2},\Delta y_{j-1},\Delta y_{j},\Delta y_{j+1},\Delta y_{j+2},v^{old}_{i,j+1}\right)\\
		  &+ D^{Ty}_{i,j+1},\\
	T^{explicit}_{i,j} =& 0.
    \label{pl31}
\end{split}
\end{equation}
The diffusion coefficients are:
\begin{equation}
    D^{Tx}_{i,j} = C^{T1}_f . \Gamma ^{\lambda  old}\!\!\mid_{x^f_i} \frac{\Delta y_j}{0.5 (\Delta x_i + \Delta x_{i-1})},\ D^{Ty}_{i,j} = C^{T1}_f . \Gamma ^{\lambda  old}\!\!\mid_{y^f_j} \frac{\Delta x_i}{0.5 (\Delta y_j + \Delta y_{j-1})}
    \label{pl32}
\end{equation}
A harmonic average between two neighboring nodes is used to calculate $\Gamma ^{\lambda  old}\!\!\mid_{x^f_i}$ and $\Gamma ^{\lambda  old}\!\!\mid_{y^f_j}$:
\begin{equation}
    \Gamma ^{\lambda old} \!\!\mid_{x^f_i} = \frac{(\Delta x_{i-1} + \Delta x_i)\Gamma^{\lambda old}_{i-1,j} \Gamma^{\lambda old}_{i,j}}{\Delta x_{i-1} \Gamma^{\lambda old}_{i,j} + \Delta x_i \Gamma^{\lambda old}_{i-1,j}},\ \Gamma ^{\lambda old} \!\!\mid_{y^f_j} = \frac{(\Delta y_{j-1} + \Delta y_j)\Gamma^{\lambda old}_{i,j-1} \Gamma^{\lambda old}_{i,j}}{\Delta y_{j-1} \Gamma^{\lambda old}_{i,j} + \Delta y_j \Gamma^{\lambda old}_{i,j-1}}
    \label{pl33}
\end{equation}
The finite-difference representation of the source term $S^T$ is expressed as:
\begin{equation}
    S^T = S^T_c + S^T_p T_{i,j},
    \label{pl28}
\end{equation}
where:
\begin{equation}
\begin{split}
    S^T_p =& 0 \\
    S^T_{c_{i,j}} =& C^{T2}_f . \Gamma^{old}_{i,j} \left\{ 2 \left[\left(\frac{u^{old}_{i+1,j} - u^{old}_{i,j}}{\Delta x_i}\right)^2 + \left(\frac{v^{old}_{i,j+1} - v^{old}_{i,j}}{\Delta y_j}\right)^2 \right] \right. \\
    & \ \ \ \ \ \ \ \ \ \ \ \ \ + \left(\frac{v^{old}(x^f_{i+1},y^v_j) - v^{old}(x^f_i,y^v_j)}{\Delta x_i} + \frac{u^{old}(x^v_i,y^f_{j+1}) - u^{old}(x^v_i,y^f_j)}{\Delta y_j}\right)^2 \\
    & \left. \ \ \ \ \ \ \ \ \ \ \ \ \ -\frac{2}{3}\left(\frac{u^{old}_{i+1,j} - u^{old}_{i,j}}{\Delta x_i} + \frac{v^{old}_{i,j+1} - v^{old}_{i,j}}{\Delta y_j}\right)^2 \right\} \Delta x_i \Delta y_j \\
    &+ C^{T3}_f p^{old}_{i,j} \left(\frac{u^{old}_{i+1,j} - u^{old}_{i,j}}{\Delta x_i} + \frac{v^{old}_{i,j+1} - v^{old}_{i,j}}{\Delta y_j}\right) \Delta x_i \Delta y_j
    \label{pl29}
\end{split}
\end{equation}
To interpolate velocities $u^{old}(x^v_i,y^f_{j+1}), u^{old}(x^v_i,y^f_j), v^{old}(x^f_{i+1},y^v_j)$ and $v^{old}(x^f_i,y^v_j)$ a bilinear interpolation between four neighboring nodes is used for each one.\\\indent
The coefficients that correspond to explicit approximation scheme of convective terms for temperature are:
\begin{equation}
\begin{split}
    a^T_0 =& \Delta t \left(a^T_1 + a^T_2 + a^T_3 + a^T_4\right) + \rho^{old}_{i,j} \Delta x_i \Delta y_j \\
    a^T_1 =& D^{Tx}_{i,j},\ a^T_2 = D^{Tx}_{i+1,j},\ a^T_3 = D^{Ty}_{i,j},\ a^T_4 = D^{Ty}_{i,j+1},\\
	T^{explicit}_{i,j} =& - F^x_{i+1,j} \left[ \text{upwind}(T^{n-1}_{i,j}, T^{n-1}_{i+1,j}, u^{n-1}_{i+1,j}) \right.\\
	&\left.\ \ \ \ \ \ \ + (T^{n-1}_{i+1,j} - T^{n-1}_{i,j}) \psi_s \left(T^{n-1}_{i-1,j},T^{n-1}_{i,j},T^{n-1}_{i+1,j},T^{n-1}_{i+2,j},\Delta x_{i-1},\Delta x_{i},\Delta x_{i+1},\Delta x_{i+2},u^{n-1}_{i+1,j}\right)\right]\\
	& + F^x_{i,j} \left[ \text{upwind}(T^{n-1}_{i-1,j}, T^{n-1}_{i,j}, u^{n-1}_{i,j}) \right.\\
	&\left.\ \ \ \ \ \ \ + (T^{n-1}_{i,j} - T^{n-1}_{i-1,j}) \psi_s \left(T^{n-1}_{i-2,j},T^{n-1}_{i-1,j},T^{n-1}_{i,j},T^{n-1}_{i+1,j},\Delta x_{i-2},\Delta x_{i-1},\Delta x_{i},\Delta x_{i+1},u^{n-1}_{i,j}\right)\right]\\
	& - F^y_{i,j+1} \left[ \text{upwind}(T^{n-1}_{i,j}, T^{n-1}_{i,j+1}, v^{n-1}_{i,j+1}) \right.\\
	&\left.\ \ \ \ \ \ \ + (T^{n-1}_{i,j+1} - T^{n-1}_{i,j}) \psi_s \left(T^{n-1}_{i,j-1},T^{n-1}_{i,j},T^{n-1}_{i,j+1},T^{n-1}_{i,j+2},\Delta y_{j-1},\Delta y_{j},\Delta y_{j+1},\Delta y_{j+2},v^{n-1}_{i,j+1}\right)\right]\\
	& + F^y_{i,j} \left[ \text{upwind}(T^{n-1}_{i,j-1}, T^{n-1}_{i,j}, v^{n-1}_{i,j}) \right.\\
	&\left.\ \ \ \ \ \ \ + (T^{n-1}_{i,j} - T^{n-1}_{i,j-1}) \psi_s \left(T^{n-1}_{i,j-2},T^{n-1}_{i,j-1},T^{n-1}_{i,j},T^{n-1}_{i,j+1},\Delta y_{j-2},\Delta y_{j-1},\Delta y_{j},\Delta y_{j+1},v^{n-1}_{i,j}\right)\right].
    \label{pl31_1}
\end{split}
\end{equation}
The sequence of calculation of numerical equations in GPU can be determined after detailed analysis of calculations in a loop along the y-axis. The numerical equations for $u$, $v$, $p$ and $T$ are (\ref{pl18}), (\ref{pl9}), (\ref{pl23}) and (\ref{pl30}), respectively. The equations for temperature (\ref{pl30}) and pseudo velocities ($\hat{u}$ (\ref{pl20}) and $\hat{v}$ (\ref{pl21})) depend on the values from the previous iteration and the previous time step that are constant variables for the iteration (loop 2). On the other hand the equations for $u$ (\ref{pl18}), $v$ (\ref{pl19}) and $p$ (\ref{pl23}) contain information from current loop 2, (see Fig. \ref{SIMPLE-TS_for_CPU_and_GPU_implicit}). The basic numerical equations can be represented as function of constant and calculated variables in loop along the y-axis as follow:
\begin{equation}
\begin{split}
	T_{i,j}=&f(\text{constant variables in loop 2})
	\label{pl29_1}
\end{split}
\end{equation}
\begin{equation}
\begin{split}
	\hat u_{i,j}=&f(\text{constant variables in loop 2})
	\label{pl29_2}
\end{split}
\end{equation}
\begin{equation}
\begin{split}
	d^u_{i,j}=&f(\text{constant variables in loop 2})
	\label{pl29_3}
\end{split}
\end{equation}
\begin{equation}
\begin{split}
	\hat v_{i,j}=&f(\text{constant variables in loop 2})
	\label{pl29_4}
\end{split}
\end{equation}
\begin{equation}
\begin{split}
	d^v_{i,j}=&f(\text{constant variables in loop 2})
	\label{pl29_5}
\end{split}
\end{equation}
\begin{equation}
\begin{split}
	p_{i,j}=&f(\hat u_{i,j}, d^u_{i,j}, \hat u_{i+1,j}, d^u_{i+1,j}, \hat v_{i,j}, d^v_{i,j}, \hat v_{i,j+1}, d^v_{i,j+1}, T_{i,j}, \\
	&\text{constant variables in loop 2})
	\label{pl29_6}
\end{split}
\end{equation}
\begin{equation}
\begin{split}
	u_{i,j}=&f(\hat u_{i,j}, d^u_{i,j}, p_{i-1,j}, p_{i,j}, \text{constant variables in loop 2})
	\label{pl29_7}
\end{split}
\end{equation}
\begin{equation}
\begin{split}
	v_{i,j}=&f(\hat v_{i,j}, d^v_{i,j}, p_{i,j-1}, p_{i,j}, \text{constant variables in loop 2})
	\label{pl29_8}
\end{split}
\end{equation}
The variables $T$, $\hat u$, $d^u$, $\hat v$ and $d^v$ depend on the constant variables in loop 2 only, see equations (\ref{pl29_1}) - (\ref{pl29_5}), therefore, they are sequence independent in the loop along the y-axis. On the other hand, $p$, $u$ and $v$ are coupled and depends on calculated variables in loop 2. In CPU algorithm, calculated variables $\hat u$, $d^u$, $\hat v$ and $d^v$ are stored in arrays in RAM memory and used in pressure equation. After pressure calculation $\hat u$, $d^u$, $\hat v$, $d^v$ and obtained pressure are used to calculate $u$ and $v$, see Fig. \ref{SIMPLE-TS_for_CPU_and_GPU_implicit}. In GPU algorithm we can reduce write/read to/from global GPU memory using local memory. The variables $p_{i-1,j}, p_{i,j-1}, p_{i,j}, \hat u_{i,j}, d^u_{i,j}, \hat u_{i+1,j}, d^u_{i+1,j}, \hat v_{i,j}, d^v_{i,j}, \hat v_{i,j+1}$ and $d^v_{i,j+1}$ are temporarily stored in local memory while $T_{i,j}$ is temporarily stored in private memory. Fig. \ref{GPU_dependent_variables_calculation} shows detailed information about subdomain calculations. In upper part Fig.\ref{GPU_dependent_variables_calculation} shows subdomains with halo regions calculated from CU and direction of internal loop along the y-axes. In blue color are calculated variables that are copied in global memory and do not have relation to current calculations. In black color are variables that will be calculated in the next iterations along the y-axis. In red color are variables calculated in previous iterations ($j-1$) along the y-axis. The red colored variables used in current iteration ($j$) are: $p_{i,j-1}, \hat v_{i,j}$ and $d^v_{i,j}$. In green color are variables calculated in current iteration along the y-axes (j): $p_{i,j}, u_{i,j}, \hat u_{i,j}, d^u_{i,j}, v_{i,j}, \hat v_{i,j+1}$ and $d^v_{i,j+1}$. In the middle part are placed relation between x-coordinate and indexing of the arrays defined in global and local device memory. In lower part Fig. \ref{GPU_dependent_variables_calculation} shows control volumes for pressure and velocities.  The detailed notations of variables together (lower left part) and separately (lower right part) that contains only calculated non-constant variables in numerical equation. One can find a pseudo code of the kernel that calculate loop 2 in Appendix A. The reuse of temporary variables stored in private and local memory decrease write/read to/from global memory, increase performance and decrease global memory requirements.\\\indent
\FloatBarrier
The organization of arrays in local memory is relatively simple. We will examine the variables $p_{i,j}$ and $p_{i,j-1}$. They was stored in array \texttt{p\_local} in the local memory. The size of the array is \texttt{Nx\_local x Ny\_p\_local}, where $\texttt{Nx\_p\_local}$ varies according to local memory size (here is fixed to 252), the number of rows $\texttt{Ny\_p\_local}$ is fixed to 2 and do not depend on local memory size. The organization of indexing of arrays is done using macros, see Fig. \ref{p_local_macros_index}. $\hat u$, $d^u$, $\hat v$ and $d^v$ are other variables stored in arrays in local memory. The arrays sizes are \texttt{Nx\_local x Ny\_u\_pseudo\_local}, \texttt{Nx\_local x Ny\_du\_local}, \texttt{Nx\_local x Ny\_v\_pseudo\_local} and \texttt{Nx\_local x Ny\_dv\_local}, respectively, where $\texttt{Ny\_u\_pseudo\_local}=1$, $\texttt{Ny\_du\_local}=1$, $\texttt{Ny\_v\_pseudo\_local}=2$ and $\texttt{Ny\_dv\_local}=2$.
\begin{figure}[htb!]
\centering
\begin{verbatim}
#define p_local(i,j) p_local_store[(i) - i0_global + halo_I_1 \
                                   + ((j) % Ny_p_local) * Nx_local]
\end{verbatim}
\caption{OpenCL pseudo code of expression implemented as macros for indexing array \texttt{p\_local\_store} defined in local memory that store data for $p_{i,j}$ and $p_{i,j-1}$, where \texttt{i0\_global} ($i^{global}_{0}$) is global index of first element in the subdomain, \texttt{halo\_I\_1} is the number of halo elements in direction \texttt{i-1}, \texttt{Nx\_local} is the number of elements along the x-axis, \texttt{Ny\_p\_local} is the number of elements along the y-axis.}
\label{p_local_macros_index}
\end{figure}
\FloatBarrier
\subsection{GPU specifics}\label{GPU_specifics_section}
As far the GPU algorithm and organization is presented mainly from the algorithmic and numerical equations point of view. The GPU's specifics are very important part of kernel development. In this and next subsection are presented GPU's specifics that influence over kernel development and performance.\\\indent
In the last few years, the performance of Graphics Processing Units (GPU's) overcame significantly the performance of Central Processor Units (CPUs). The reason is that desktop CPUs have 4 cores, while the GPU's have much more Compute Units (CU). As an example, AMD Radeon R9 280X have 32 CU. The CU are very similar to CPU core; therefore AMD Radeon R9 280X possess 8 times more "cores" than ordinary desktop CPU. GPU architecture is very similar to CPU one: GPU CU corresponds to CPU core; GPU private memory corresponds to CPU registers; GPU local memory corresponds to CPU L1 cache, and GPU global memory corresponds to computer RAM memory. On the other hand CPU core handle one thread, while CU can handle a lots of threads (work-items) simultaneously. One CU of AMD Radeon R9 280X can handle maximum 2560 work items. Unfortunately, the block of work-items executed together are 64 (a wavefront) \cite{AMD_Parallel_OpenCL_November2013}. Up to four work-items from the same wavefront on the same stream core are pipelined to hide latency due to memory accesses and processing element operations. Therefore one CU maximum pipeline is 4x64=256 wave-items, see \cite{AMD_Parallel_OpenCL_November2013}.\\\indent
On the other hand, GPU's are specific devices for parallel calculations. GPU's possess smaller private (registers), local and global memory compared to CPUs. Therefore, the suitable algorithm should possess excellent parallel scalability and to be highly arithmetic intensive i.e. to do very intensive calculations over relatively small number of variables.\\\indent
GPU devices require highly arithmetic intensive algorithms to reach good performance. Volkov present importance of registers usage and instruction-level parallelism (ILP) to reach better performance at lower occupancy \cite{volkov2010}. This idea was adopted and implemented appropriately in presented GPU algorithm. On the other hand, a small expressions can be optimized very precisely, while numerical equations of CFD are very complex and large. Independent variables for 2D case are 4 ($p$, $T$, $u$ and $v$) and requires up to 388 floating point operations per numerical equation. For each node, we have to copy from memory values from 15 to 20 control volumes. The number of all copied variables per node is approximately from 60 to 80. The number floating point operations per node is from 687 to 1229, see Table \ref{N_FLOP_GPU_explicit} and Table \ref{N_FLOP_GPU_implicit}. Therefore, the arithmetic intensity is high: from $687/80 \approx 9$ to $1229/60 \approx 20$. Manual optimization of these big expressions is hard work, where many errors can occur. Furthermore, the changes in partial differential equations or numerical scheme will require corresponding changes in a code. Almost all optimization was left to a compiler to overcome difficulties related to manual optimization. The compiler organizes the copy of data from global to private memory and reuse of calculated numerical expressions in a code. The basic elements in numerical expressions were substituted using macros. The number of floating point operations was count after macros expansion in expressions, Table \ref{N_FLOP_GPU_explicit} and Table \ref{N_FLOP_GPU_implicit}. This number of floating point operations was used to calculate GPU performance. The compiler can reduce number of floating point operations using common subexpression elimination (CSE), but analysis of assembler is out of the scope of this paper, and CSE reduction was not taken into account. Presented idea in this paper is to use big expressions that gives possibility for ILP, use registers to store temporary variables and leave copy of data between memories and latency hiding organization to a compiler.
\FloatBarrier
\subsection{Tips and tricks}\label{GPU_tips_section}
Execution of some operations on GPU's cause great performance lost. To overcome this have to use alternative operation or code reorganization. The execution time of division and square root are slow operations. To speedup code density ($\rho = p / T$) and diffusion coefficient ($\Gamma = \sqrt{T}$) was stored in global memory. The logical operations could use with care in GPU code. Even a couple logical operations per kernel could decrease performance a couple of times. One can replace logical operators with equivalent multiplication and addition operations, taking into account specifics of data converting between boolean and integer variables. The boolean type, known in C++ as \verb"bool", can only represent one of two states, true or false. Boolean type true and false converted to an integer type, known in C++ as \verb"int", are 1 and 0, respectively. The conversion of integer type to boolean type is 0 to false and every value different from 0 to true; therefore -1, 1, -10, 10 are true. The conversion between boolean type and floating-point types is the same. Take into account conversion between data types logical AND can be replaced with multiplication, logical OR can be replaced with addition, see Table \ref{Logical_operators_equivalent_table}. The power function also could decrease code performance. In presented algorithm only values to the power of 2 were used and were substituted with multiplication, see Table \ref{Logical_operators_equivalent_table}. After all the slow operations as logical AND, logical OR and power function can be replaced with faster equivalent operations that contain the fastest operations as addition, subtraction, multiplication and type conversion.\\\indent
Simplified test code could make the analysis of performance much easier and show some important tendencies. Testing of corresponding operations directly in CFD code could require additional changes and the results could be unclear. The simplified test code used by the author was the equation of temperature. All variables were defined in private memory. Simplified temperature equation was calculated within a loop, where variables depend only on a counter and no variables were copied to/from global memory. The simplified test code shows performance lost unquestionably when use division, logical AND, logical OR and power function. Simplified test code shows excellent performance of equivalent operations in Table \ref{Logical_operators_equivalent_table}. The developer could check every operator performance when substitute it with multiplication or addition. This substitution will give the information about the performance of operator according the fastest operations as addition and multiplication. If the code accelerates more than two times, it is worth to try to find faster equivalent.\\\indent
The dependence of the code performance from specific operators is expected, because the GPU's development was boosted initially by the game industry, where for graphical processing are calculate relatively simple expressions. Furthermore, the variables are integer or floating point with single precision accuracy. On the other hand, scientific expressions could be big and complicated expressions that include different operators and double precision floating point operations. The GPU's are using actively for a scientific computing only for a couple of years and hardware, drivers and development tools are in a process of rapid development. Nevertheless, short period the fastest supercomputers at the moment use GPU's as coprocessors, \cite{SuperkompiutriTop500}.\\\indent
\begin{table}[htb!]
\begin{center}
\begin{tabular}{c c c}
\toprule
  \textbf{Operation}
& \textbf{C++ operation}
& \textbf{Faster equivalent for GPU} \\
\toprule
Logical AND & $x\ \&\&\ y$ & $x * y$\\
\midrule
Logical OR  & $x\ ||\ y$   & $(bool)(x + y)$\\
\midrule
Logical NOT & $!x$         & $(1 - (bool)(x))$\\
\midrule
Logical XOR & - & $(bool)(x - y)$\\
\midrule
x to the power of 2 & $pow(x,2)$ or $pown(x,2)$ & $x * x$ \\
\bottomrule
\end{tabular}
\caption{Equivalent operations using multiplication and addition for faster calculations on GPU.}
\label{Logical_operators_equivalent_table}
\end{center}
\end{table}
Other specific in GPU development is related to if-then-else conditions. If-then-else conditions are widely used in CPU codes to reduce calculations that speedup the code after all. A conditional of the form if-then-else generates branching (code serialization) in GPU codes and could slow down the calculations significantly. According AMD Accelerated Parallel Processing OpenCL Programming Guide, \cite{AMD_Parallel_OpenCL_November2013} page 2-4, branching occurs when: "If work-items within a wavefront diverge, all paths are executed serially." This serialization could be a reason for the significant performance loss of GPU code, especially when conditions are nested, \cite{AMD_Parallel_OpenCL_November2013} page 7-54: "When if blocks are nested, the results are twice as bad; in general, if blocks are nested $k$ levels deep, $2^k$ nested conditional structures are generated." For more information about branching see \cite{AMD_Parallel_OpenCL_November2013}, sections 2.1.3 Flow Control, 6.8.3 General Tips, 6.8.7 Optimizing Kernels for Southern Island GPU's and 7.10.7 Optimizing Kernels for Evergreen and 69XX-Series GPU's. The solution is to change the algorithm in a way to avoid if-then-else conditions. One of the possible approaches are Kronecker delta function or ternary operator (?:). Kronecker delta function reduces branching but increase number of operations. Density calculations in middle points using upwind scheme is a typical example of application of Kronecker delta function and ternary operator, Fig. \ref{Branching_CPU_GPU_code}. Substituting everything with Kronecker delta function significantly increase the number of floating points operations, when calculating control volumes on the walls. The control volumes on the walls require approximately two times more floating point operations than the control volumes in the fluid. The control volumes on the walls are a small part of control volumes in the computational domain. Therefore, an if-then-else condition was applied to separate control volumes that are on the wall and inside the fluid, see Fig. \ref{GPU_code_boundary_condition_check}. Kronecker delta function or ternary operator was applied to the control volumes on the walls to avoid nested if-then-else conditions. In GPU code, the performance difference between Kronecker delta function and ternary operator is around 10\% and varies according approximation scheme. After all the application of if-then-else conditions requires to take into account algorithm's and code's specifics and to do appropriate tests to understand influence of implementation of different operators over performance.\\\indent
\begin{figure}[htb!]
\begin{minipage}{0.48\textwidth}
    \centering
    \textbf{CPU code}
    \begin{verbatim}
    if(0<u(i,j))
       rho_u(i,j)=rho(i-1,j);
    else
       rho_u(i,j)=rho(i,j);
    \end{verbatim}
    \ \\
    \ \\
    \ \\
    \ \\
    \ \\
\end{minipage}
\begin{minipage}{0.48\textwidth}
    \begin{center}
      \textbf{GPU code}
    \end{center}
    Kronecker delta expansion
    \begin{verbatim}
    rho_u(i,j)=(0<u(i,j))*rho(i-1,j)
              +(!(0<u(i,j)))*rho(i,j);
    \end{verbatim}
    OR\\
    ternary operator
    \begin{verbatim}
    rho_u(i,j)=(0<u(i,j))?rho(i-1,j):rho(i,j);
    \end{verbatim}
\end{minipage}
\caption{Calculate density in middle points using upwind scheme: CPU code (left part) and GPU codes (right part).}
\label{Branching_CPU_GPU_code}
\end{figure}

\begin{figure}[htb!]
\centering
\begin{verbatim}
  if(control_volume_in_fluid(i,j))
  {
    //Calculate numerical equations
    //without implementation of boundary conditions.
  }
  else if(control_volume_on_the_wall(i,j))
  {
    //Calculate numerical equations
    //with implementation of boundary conditions.

    //The boundary conditions implemenation
    //increase number of floating point operations
    //approximately two times.
  }
\end{verbatim}
\caption{GPU pseudo code: check place of calculated control volume.}
\label{GPU_code_boundary_condition_check}
\end{figure}
\FloatBarrier

\section{Test case formulation}
\ \\[-7mm]
\begin{figure}[htb!]
    \centering
    \includegraphics[trim = 0 0 0 0,clip=true, keepaspectratio=true, width=0.7\linewidth]{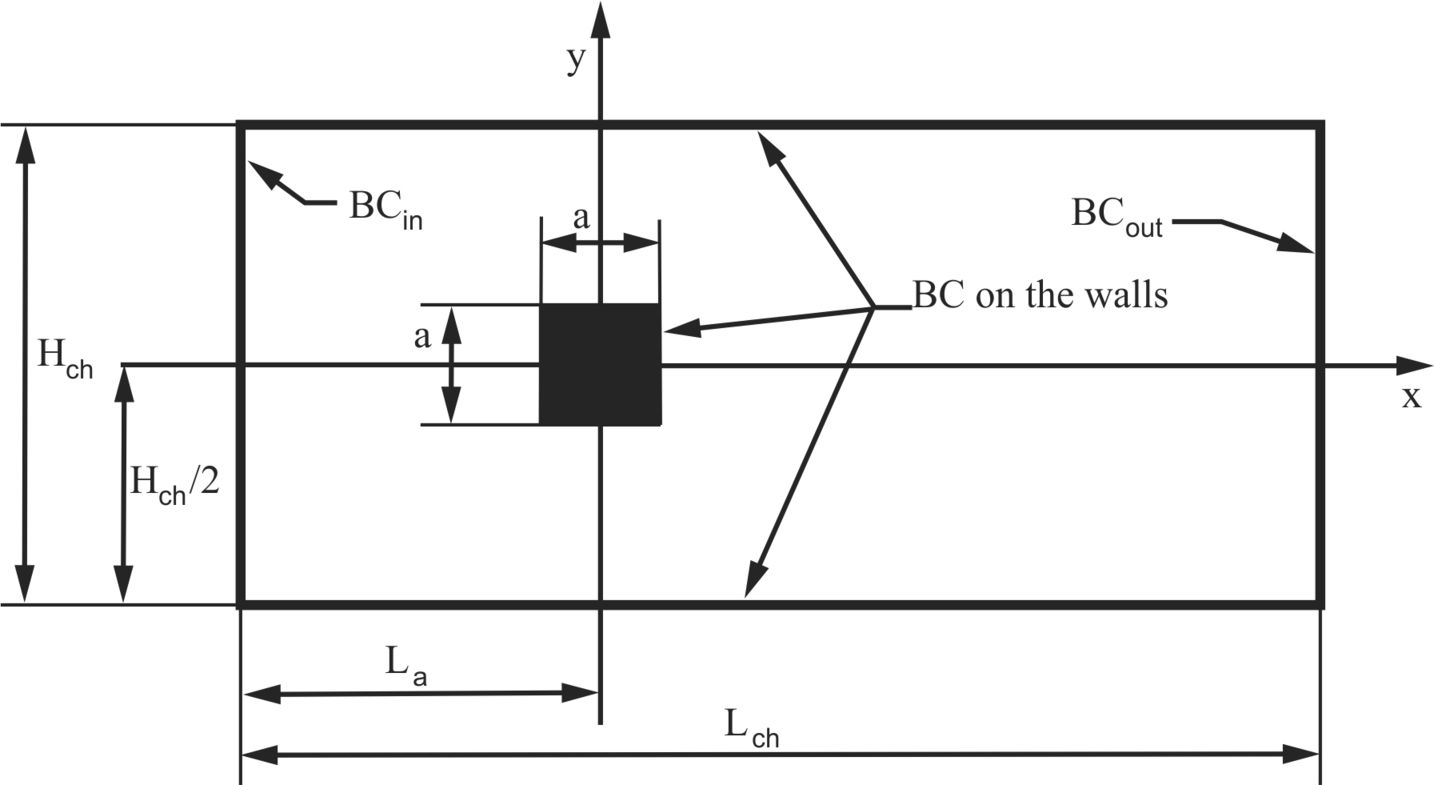}
    \caption{Flow geometry for a square-shaped particle with size $a$ confined in a channel with length $L_{ch}$ and height $H_{ch}$.}
    \label{SquareInMicrochannel}
\end{figure}
As a test case we use flow past a square particle(s) in a microchannel. The fluid model is described by the Navier-Stokes-Fourier equations (\ref{pl4}) - (\ref{pl5}). For gaseous microflow description, we use the model of a compressible, viscous hard sphere gas with diffusion coefficients determined by the first approximation of the Chapman-Enskog theory for low Knudsen numbers \cite{Stefanov2002}. The Knudsen number ($Kn$), a nondimensional parameter, determines the degree of appropriateness of the continuum model. It is defined as the ratio of mean free path $\ell_0$ to the macroscopic length scale of the physical system $L$ ($Kn = \ell_0 / L$). For the calculated case, the Knudsen number is equal to $Kn=0.001$  and the speed is equal to Mach number $M=2.43$ at the channel inlet. For a hard-sphere gas, the viscosity coefficient $\mu$ and the heat conduction coefficient $\lambda$ (first approximations are sufficient for our considerations) read  as follows:
\begin{equation}
    \mu = \mu_h \sqrt{T},\ \mu_h = (5/16) \rho_0 \ell_0 V_{th} \sqrt{\pi}
    \label{pl35}
\end{equation}
\begin{equation}
    \lambda = \lambda_h \sqrt{T},\ \lambda_h = (15/32) c_p \rho_0 \ell_0 V_{th} \sqrt{\pi}
    \label{pl36}
\end{equation}\indent
The Prandtl number is given by $Pr = 2 / 3$, $\gamma = c_p / c_v = 5 / 3$. The dimensionless system of equations (\ref{pl4}) - (\ref{pl5}) is scaled by the following reference quantities, as given in \cite{Stefanov2002}: molecular thermal velocity $V_0 = V_{th} = \sqrt{2 R T_0}$ for velocity, for length - square size $a$ (Fig. \ref{uT_fields}), for time - $t_0 = a / V_0$, the reference pressure ($p_0$) is pressure at the inflow of the channel, the reference temperature ($T_0$) is equal to the channel walls, reference density ($\rho_0$) is calculated using equation of state (\ref{pl5}). The corresponding non-dimensional parameters in the equation system (\ref{pl4}) - (\ref{pl5}) are computed by using the following formulas:
\begin{equation}
\begin{split}
    &A = 0.5,\ B = \frac{5 \sqrt{\pi}}{16} Kn,\ \Gamma = \Gamma^\lambda = \sqrt{T} \\
    &C^{T1} = Kn \sqrt{\pi \frac{225}{1024}},\ C^{T2} = \frac{\sqrt{\pi}}{4} Kn,\ C^{T3} = \frac{2}{5}
    \label{pl37}
\end{split}
\end{equation}\\\indent
Fig. \ref{SquareInMicrochannel} shows the test case geometry. The channel length is $L_{ch}=201.6$, the channel inlet is $L_a=5.5$. The channel height ($H_{ch}$) varies from 10 to 200 because was investigated the influence of iterations in kernel along the y-axis over the performance. The uniform Cartesian grid with special steps $\Delta x = \Delta y = \Delta = 0.05$ was used. The problem is considered in a local Cartesian coordinate system, which is moving with the particle. Thus for an observer moving along with the particle the problem is transformed to a consideration of a gas flow past a stationary square confined in a microchannel with moving walls. Velocity-slip and temperature-jump boundary conditions \cite{Cercignani1975} are imposed on the walls of the channel and the square. The velocity-slip BC is given as:
\begin{equation}
	v_s - v_w = \zeta \left.{\frac{\partial v}{\partial n}}\right|_s,
	\label{pl38}
\end{equation}
where $v_s$ is velocity of the gas at the solid wall surface, $v_w$ is velocity of the wall, $\zeta = 1.1466 . Kn_{local} = 1.1466 . Kn / \rho_{local}$, $Kn_{local}$ is the local Knudsen number, $\rho_{local}$ is the local density, $\left.{\frac{\partial v}{\partial n}}\right|_s$ is the derivative of velocity normal to the wall surface. The temperature-jump boundary condition is:
\begin{equation}
	T_s - T_w = \tau \left.{\frac{\partial T}{\partial n}}\right|_s,
	\label{pl39}
\end{equation}\\
where $T_s$ is temperature of the gas at the wall surface , $T_w$ is temperature of the wall, $\tau = 2.1904 . Kn_{local} = 2.1904 . Kn / \rho_{local}$, $\left.{\frac{\partial T}{\partial n}}\right|_s$ is the derivative of temperature normal to the wall surface.
\FloatBarrier

\section{Speedup analysis} \label{Speedup_analysis}
\begin{figure}[htb!]
    \centering
    \includegraphics[keepaspectratio=true, width=0.49\textwidth]{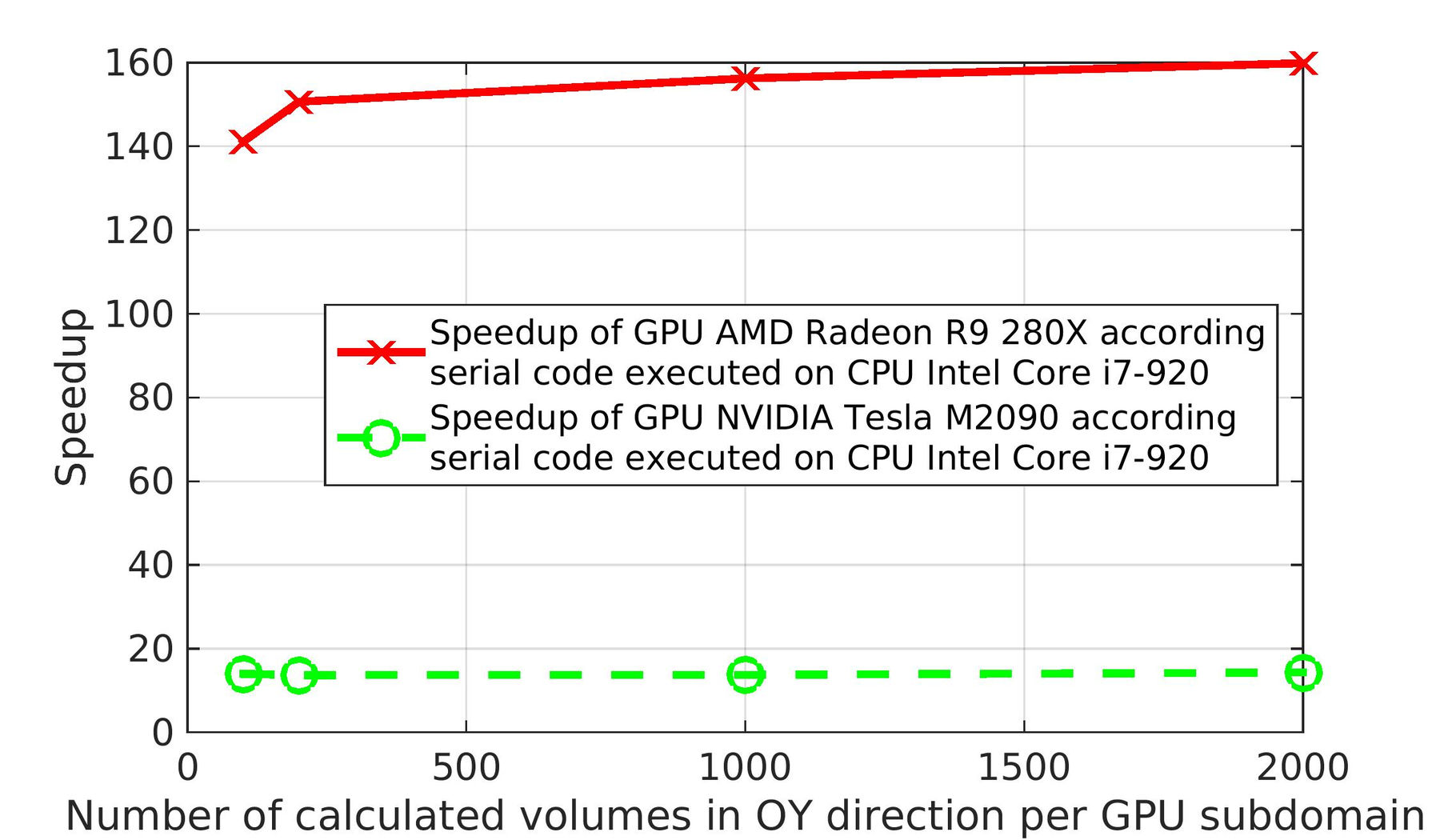}
    \ \includegraphics[keepaspectratio=true, width=0.49\textwidth]{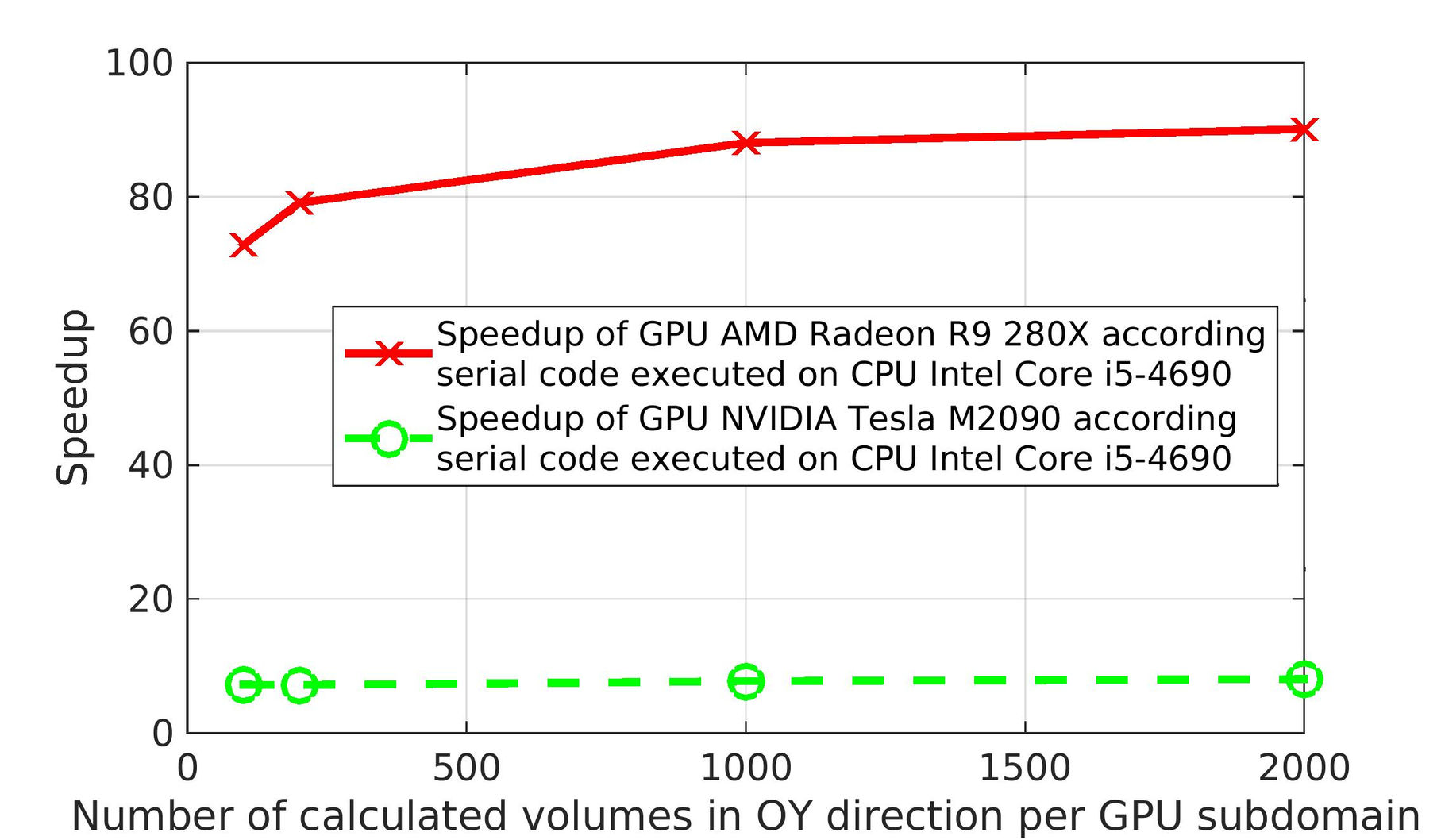}
    \caption{The speedup was obtained by comparison two GPU's: GPU AMD Radeon R9 280X and GPU NVIDIA Tesla M2090 with serial code executed on two CPUs: CPU Intel Core i7-920 (left part) and CPU Intel Core i5-4690 (right part). In this test case, explicit TVD second-order scheme with Van-Leer limiter approximates convective terms.}
    \label{Speedup_explicit_TVD_GPU_to_CPU}
\end{figure}

\begin{figure}[htb!]
    \centering
    \includegraphics[keepaspectratio=true, width=0.49\textwidth]{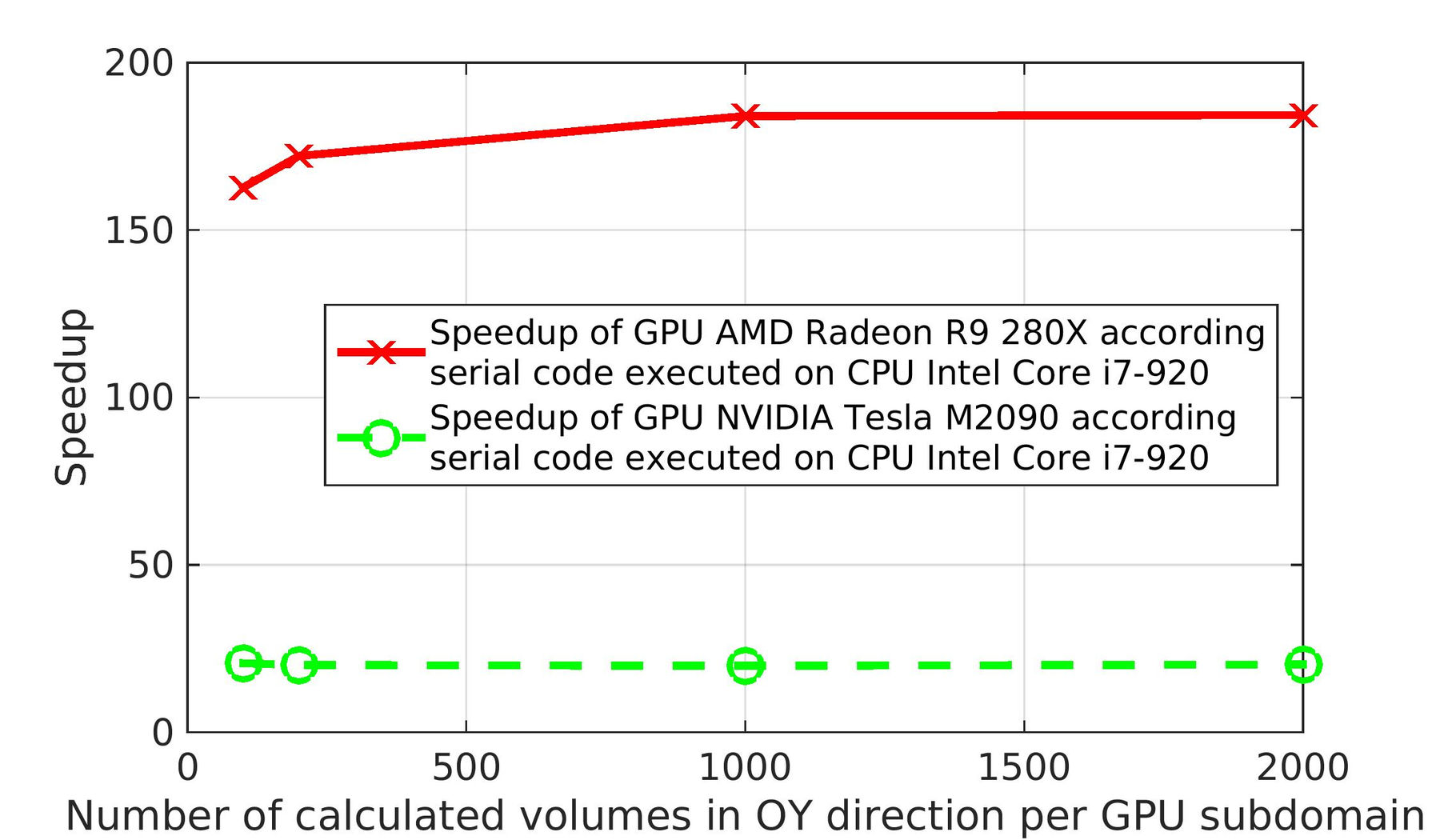}
    \ \includegraphics[keepaspectratio=true, width=0.49\textwidth]{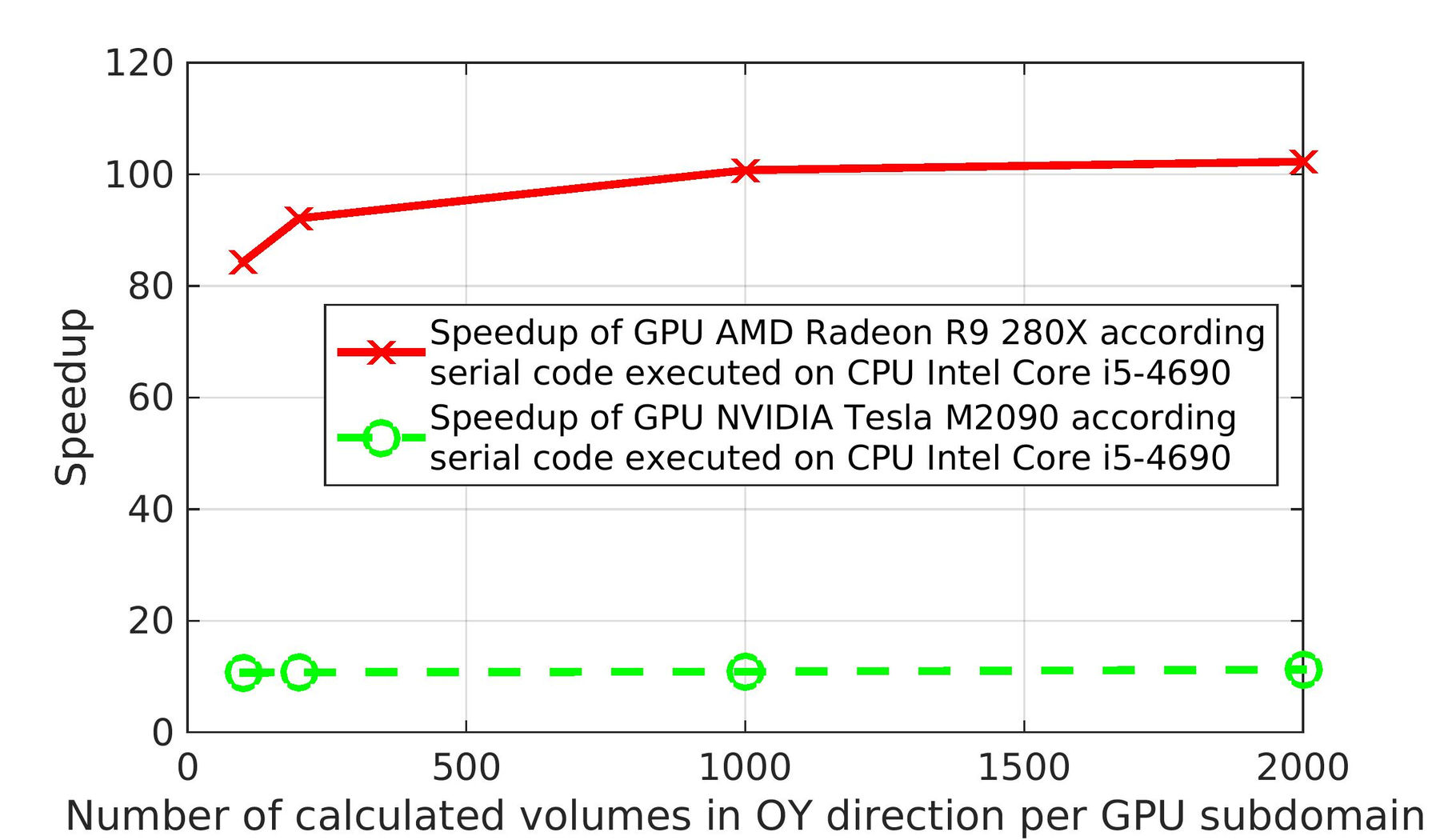}
    \caption{The speedup was obtained by comparison two GPU's: GPU AMD Radeon R9 280X and GPU NVIDIA Tesla M2090 with serial code executed on two CPUs: CPU Intel Core i7-920 (left part) and CPU Intel Core i5-4690 (right part). In this test case, explicit upwind 1-st order scheme with Van-Leer limiter approximates convective terms.}
    \label{Speedup_explicit_upwind_GPU_to_CPU}
\end{figure}

\begin{figure}[htb!]
    \centering
    \includegraphics[keepaspectratio=true, width=0.49\textwidth]{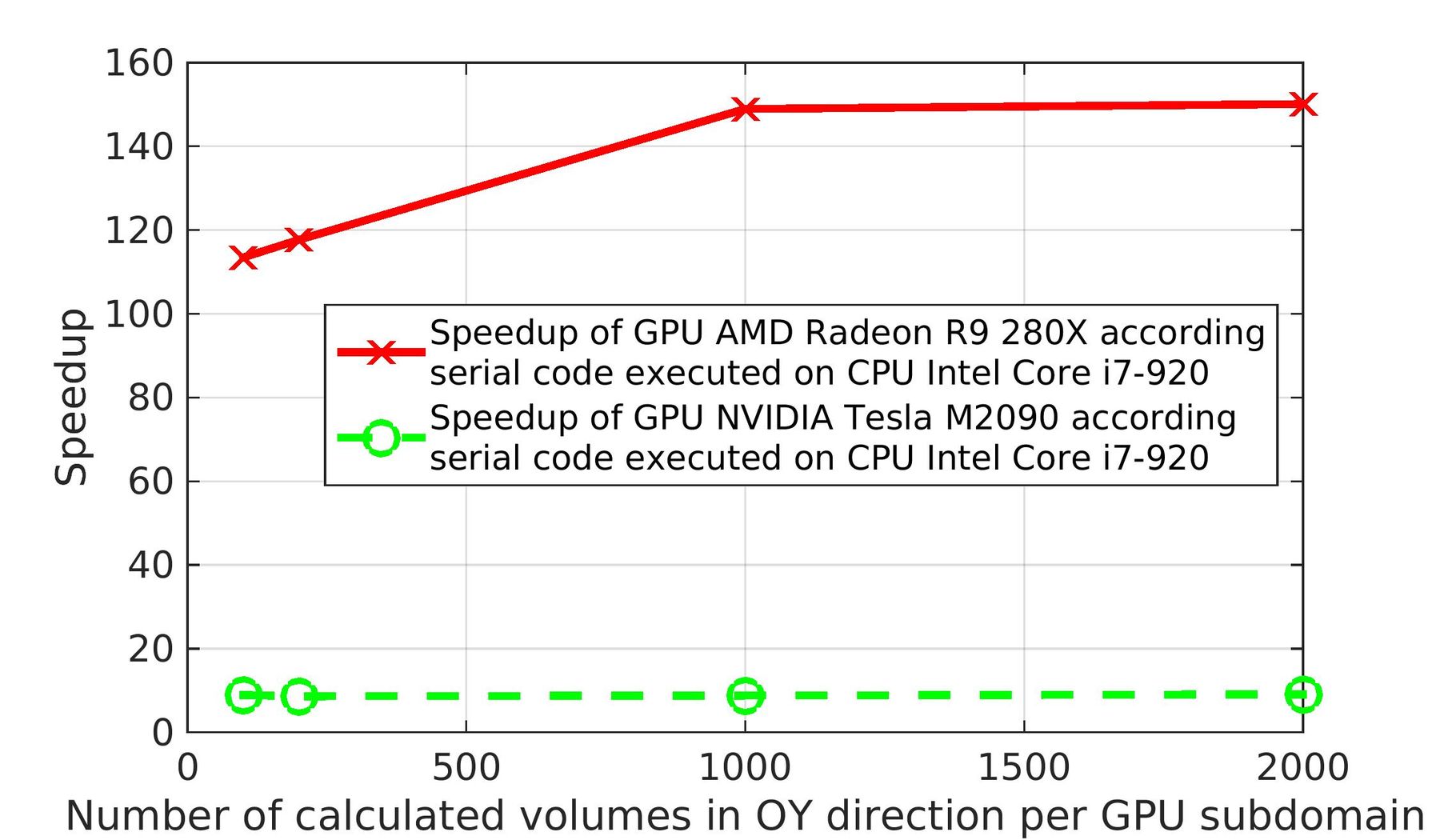}
    \ \includegraphics[keepaspectratio=true, width=0.49\textwidth]{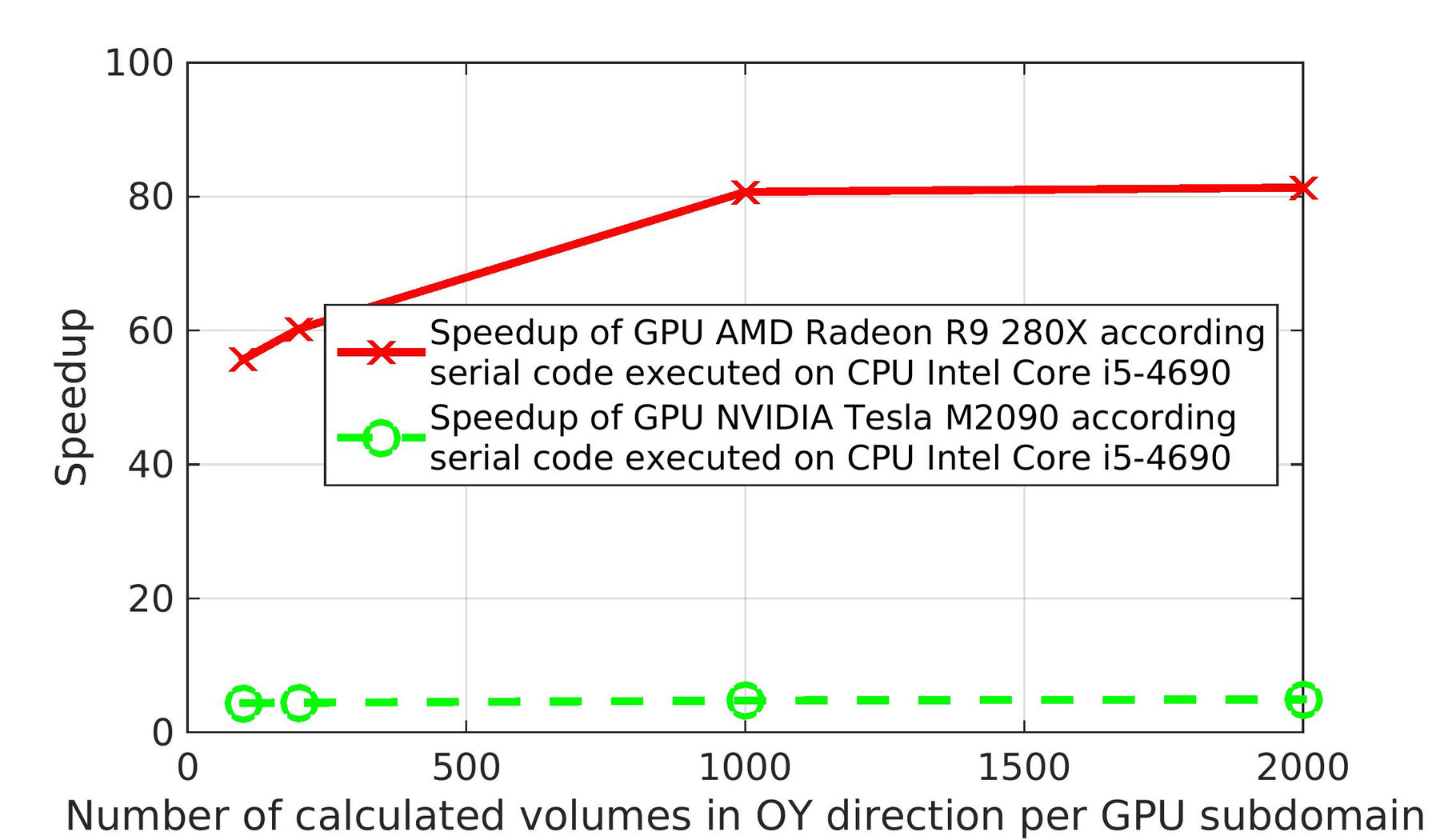}
    \caption{The speedup was obtained by comparison two GPU's: GPU AMD Radeon R9 280X and GPU NVIDIA Tesla M2090 with serial code executed on two CPUs: CPU Intel Core i7-920 (left part) and CPU Intel Core i5-4690 (right part). In this test case, implicit TVD second-order scheme with Van-Leer limiter approximates convective terms.}
    \label{Speedup_implicit_TVD_GPU_to_CPU}
\end{figure}

\begin{figure}[htb!]
    \centering
    \includegraphics[keepaspectratio=true, width=0.49\textwidth]{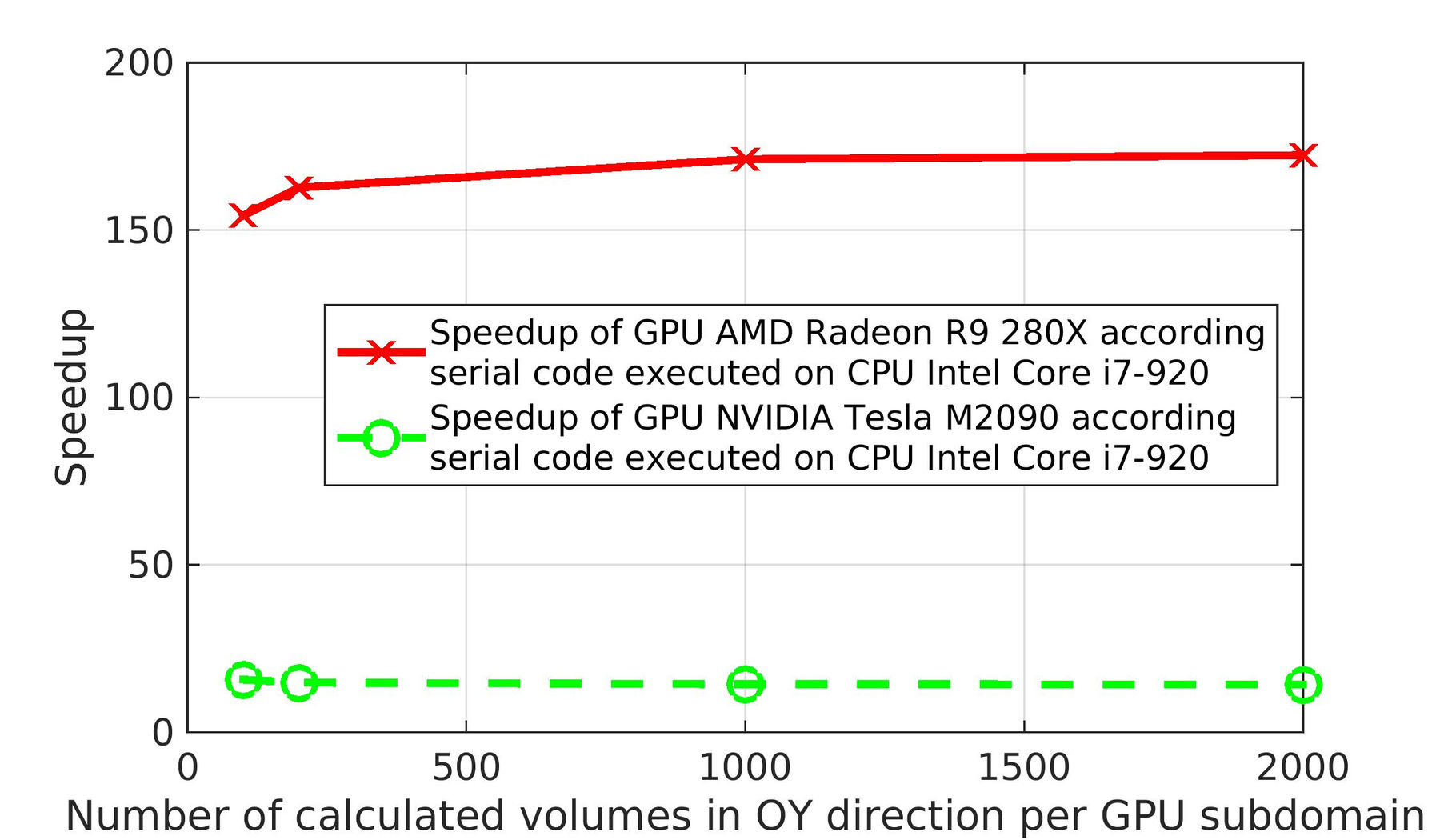}
    \ \includegraphics[keepaspectratio=true, width=0.49\textwidth]{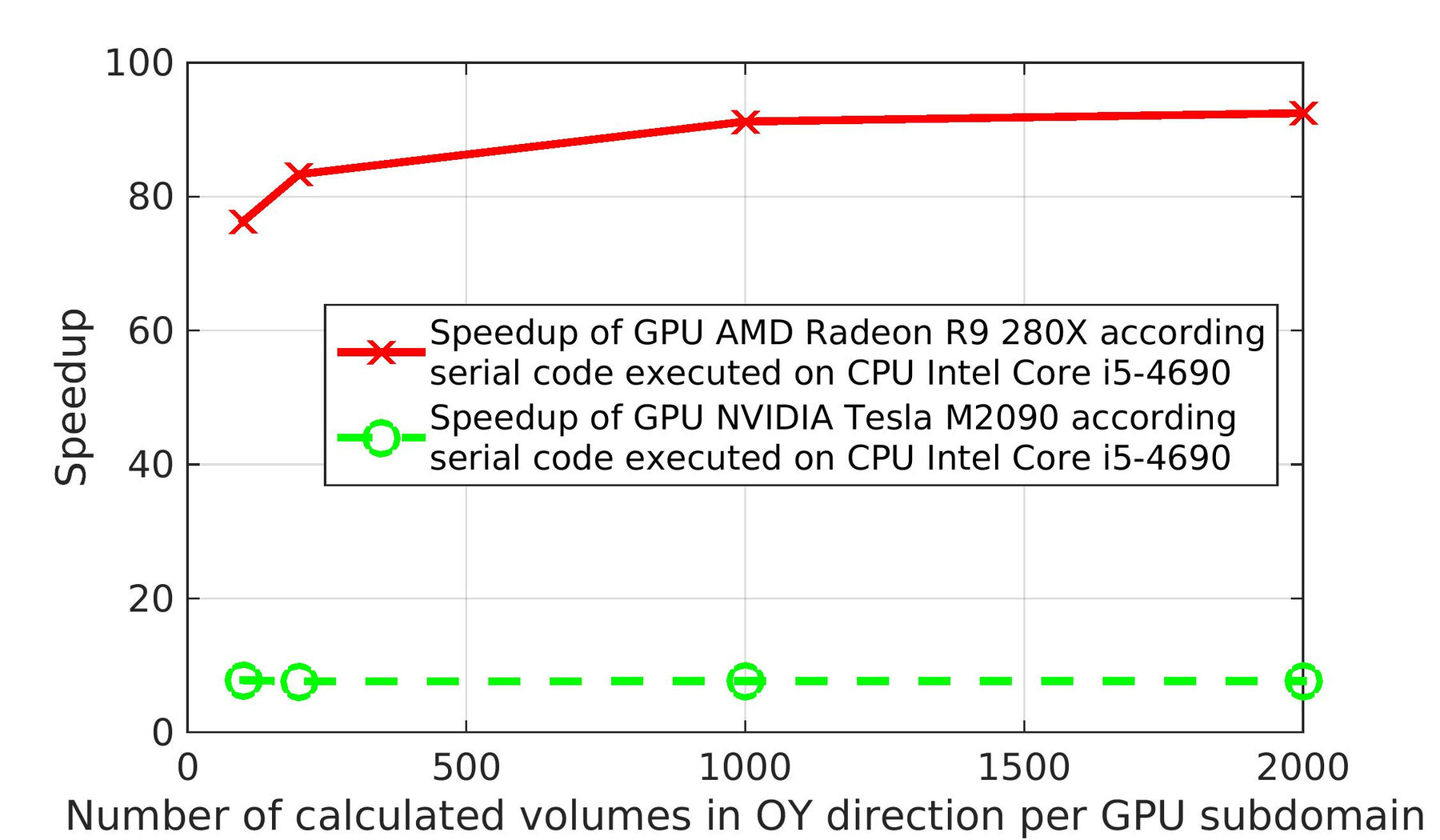}
    \caption{The speedup was obtained by comparison two GPU's: GPU AMD Radeon R9 280X and GPU NVIDIA Tesla M2090 with serial code executed on two CPUs: CPU Intel Core i7-920 (left part) and CPU Intel Core i5-4690 (right part). In this test case, implicit upwind 1-st order scheme with Van-Leer limiter approximates convective terms.}
    \label{Speedup_implicit_upwind_GPU_to_CPU}
\end{figure}

The GPU code speedup was obtained with comparison with serial CPU code. Both codes use double precision floating point operations. The GPU and CPU codes demonstrate agreement within double precision floating point accuracy that is 15 significant digits or $10^{-15}$. The GPU kernels are written in OpenCL that make it portable without modifications to AMD and NVIDIA GPU's. The CPU code is written in C++. GPU code performance was obtained on AMD Radeon R9 280X and NVIDIA Tesla M2090 while CPU code performance was tested on Intel Core i5-4690 and Intel Core i7-920. AMD Radeon R9 280X is a gaming GPU with following characteristics: release date 10.2013, peak double precision floating point performance 1024[GF/s], memory size 3072[MB], number of Compute Units 32, see \cite{AMD_GPU_wiki}. NVIDIA Tesla M2090 is a server GPU for scientific calculations with following characteristics: release date 03.2011, peak double precision floating point performance 665[GF/s], memory size 6[GB], number of Compute Units 16, see \cite{NVIDIA_GPU_M2090_Product_brief}. Intel Core i7-920 characteristics are as follow: launch date Q4'08, Instruction Set Extensions is SSE4.2, Processor Base Frequency is 2.66[GHz], Max Turbo Frequency 2.93[GHz], Number of Cores 4, Number of Threads 8, see \cite{CPU_Intel_Core_i7_920_breaf}. CPU Intel Core i5-4690 is the next generation core architecture according Intel Core i7-920. Intel Core i5-4690 characteristics are as follow: launch date Q2'14, Instruction Set Extensions is SSE4.1/4.2, AVX 2.0, Processor Base Frequency is 3.5[GHz], Max Turbo Frequency 3.9[GHz], Number of Cores 4, Number of Threads 4, see \cite{CPU_Intel_Core_i5_4690_breaf}.\\\indent
Speedup tests were obtained on tree configurations:
\begin{itemize}
  \item CPU Core i5-4690, motherboard Gigabyte Z97-HD3, dual channel memory 32[GB] at 1333[MHz], video card is Sapphire Tri-X R9 280X (AMD Radeon R9 280X), the operating system is Debian GNU/Linux 7.7 (wheezy) x64bit, C++ compiler g++ version is 4.6.3-14, video card driver version is AMD Catalyst 14.501.1003 and AMD OpenCL SDK version is 2.9-1. This configuration is part of cluster of Institute of Mechanics at the Bulgarian Academy of Sciences (IMech-BAS).
  \item CPU Core i7-920, motherboard ASUS P6TD DELUXE, triple channel memory 18[GB] at 1333[MHz], video card is GeForce GTX 260, the operating system is Debian GNU/Linux 7.7 (wheezy) x64bit and C++ compiler g++ version is 4.6.3-14.
  \item Two CPU Intel Xeon E5649, memory 96[GB], eight GPU NVIDIA Tesla M2090, the operating system is Scientific Linux release 6.6 (Carbon) x64bit, C++ compiler g++ version is 4.4.7-11, video card driver version is 319.37 and NVIDIA CUDA SDK version is 5-5. This configuration is part of HPCG cluster \cite{atanassov2014}. The HPCG cluster is located at Institute of Information and Communication Technologies at the Bulgarian Academy of Sciences (IICT-BAS).
\end{itemize}
A multiple GPU system needs appropriate cooling. One of the easiest ways to build multiple GPU systems with good cooling is to use PCI-e 1x to 16x riser cable. PCI-e 1x to 16x riser cables are about 20cm long and connect GPU with motherboard PCI-e slot. Mounting GPU on a distance of a motherboard improve cooling of GPU and make possible to increase the number of GPU's per configuration significantly. A maximum number of GPU's depends on a number of PCI Express 16x and 1x slots per motherboard. The not specialized motherboard has from 3 to 8 PCI Express slots. The main disadvantage of PCI-e 1x to 16x riser cables is slow down the bandwidth between GPU and motherboard. The performance of presented GPU algorithm does not depend on the bandwidth between GPU and motherboard. The program copy data to the device (GPU) global memory at the beginning of calculations. During the calculations program copy from device memory to host memory only maximum residuals of loop 2. That is a small quantity of information and do not need high bandwidth. A test shows that presented GPU code obtained the same performance on a system when GPU AMD Radeon R9 280X was connected to the motherboard with PCI-e 1x to 16x riser cable as the case when GPU AMD Radeon R9 280X was connected to the motherboard directly. After all PCI-e 1x to 16x riser cable improve cooling of configuration and increase the number of GPU's per motherboard without performance loss of presented algorithm.\\\indent
The memory requirements are important for GPU codes because GPU's memory is fixed and less than CPU memory. The presented GPU algorithm does an internal loop along the y-axis and uses pseudo velocities, pressure coefficients and other temporary variables only in local memory that decrease arrays in global memory. Finally, the GPU code requires two times less memory than CPU code. GPU code defines 5.9 million control volumes per 1 GB of global GPU memory for explicit and implicit schemes.\\\indent
The computational domain was decomposition to 32 subdomains, 16 subdomains along the x-axis and 2 in along the y-axis. The number of CU (Compute Units) of GPU AMD Radeon R6 280X is 32 that mean subdomain per CU. The number of CU of NVIDIA Tesla M2090 is 16 that mean two subdomains per CU. A GPU reach maximum performance when the number of subdomains is multiple to the number of CU.\\\indent
\begin{figure}[htb!]
\centering
 \includegraphics[trim=600 1320 580 1400,clip=true,width=1.0\textwidth]{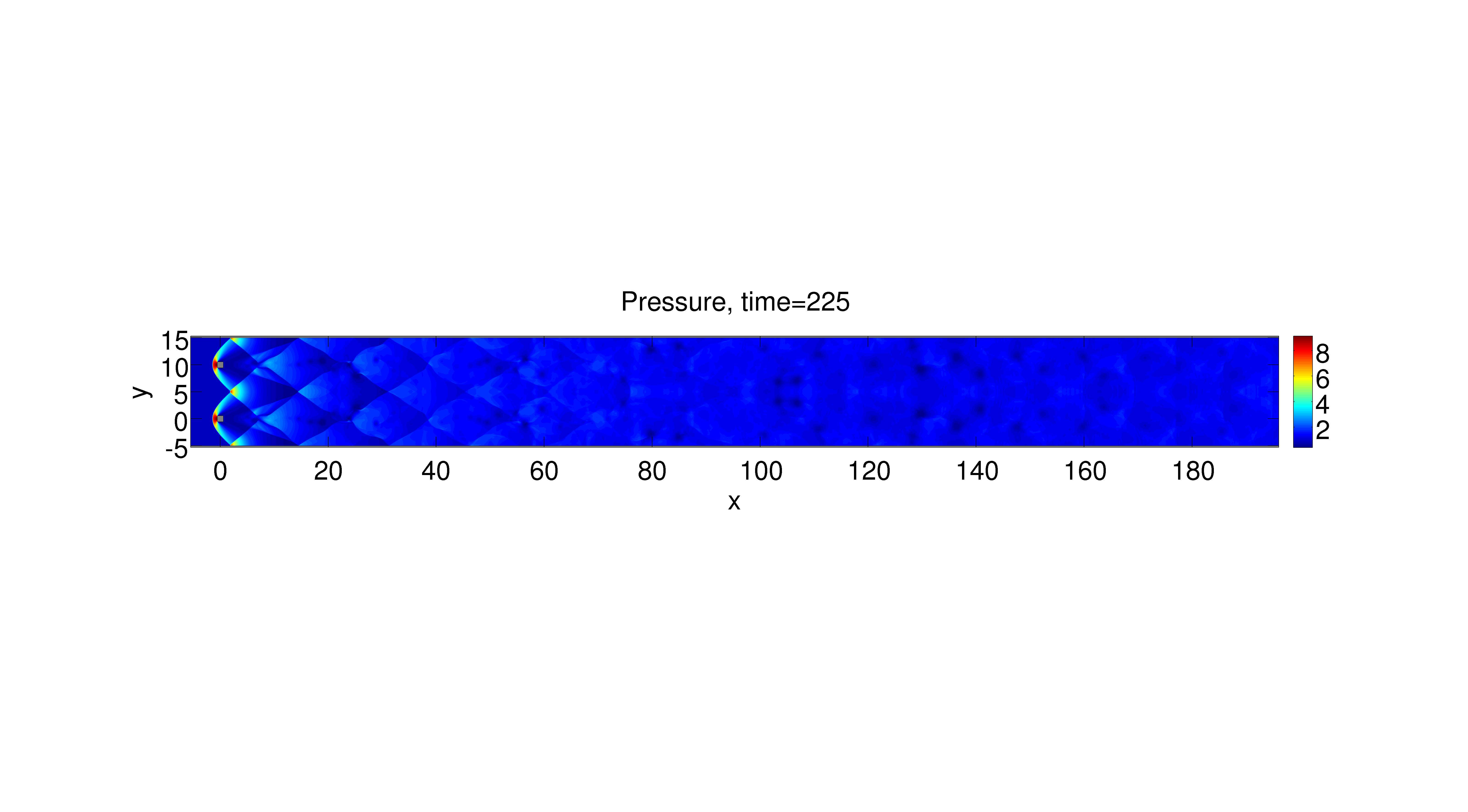}\ \\
 \includegraphics[trim=600 1320 580 1400,clip=true,width=1.0\textwidth]{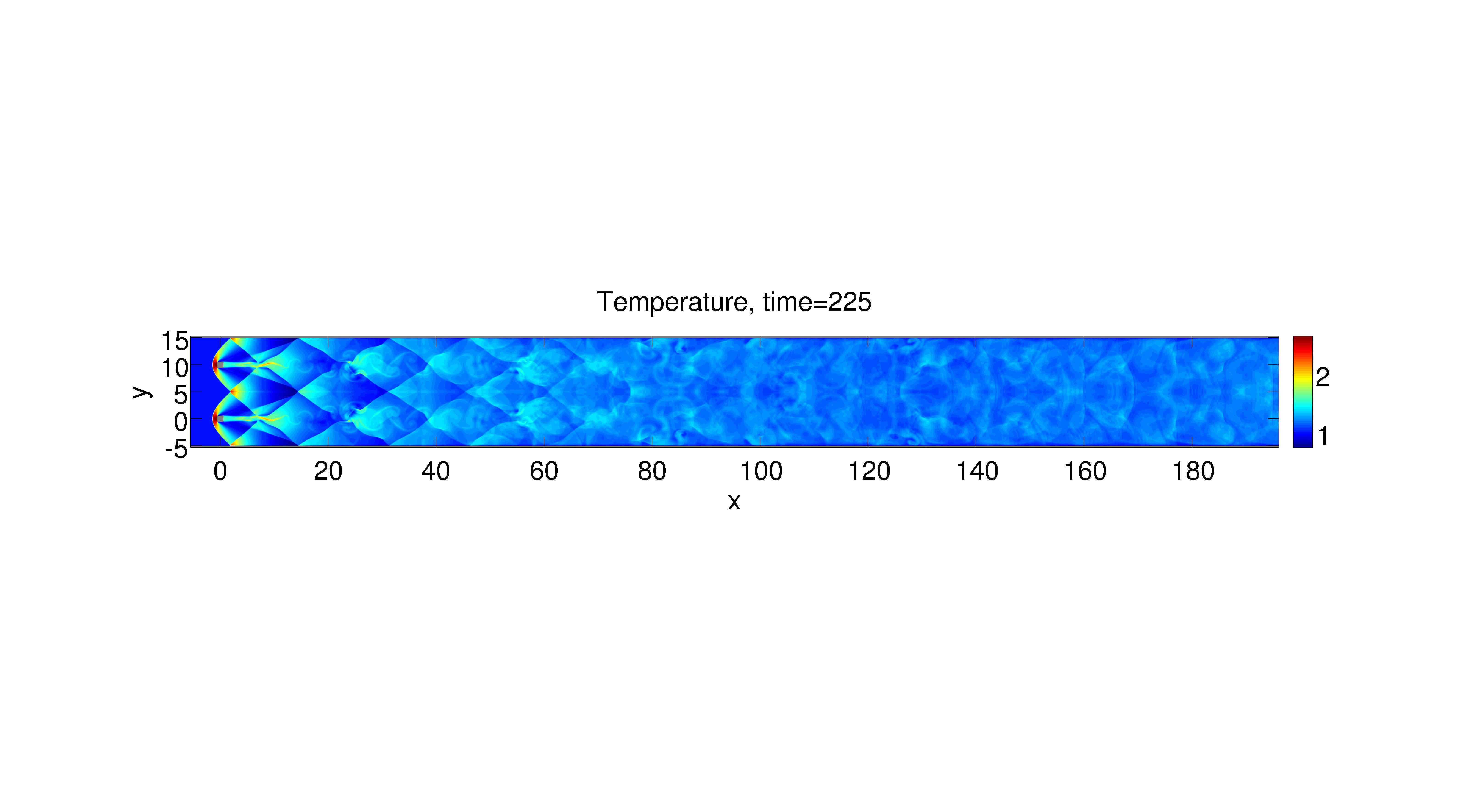}
 \includegraphics[trim=600 1320 580 1400,clip=true,width=1.0\textwidth]{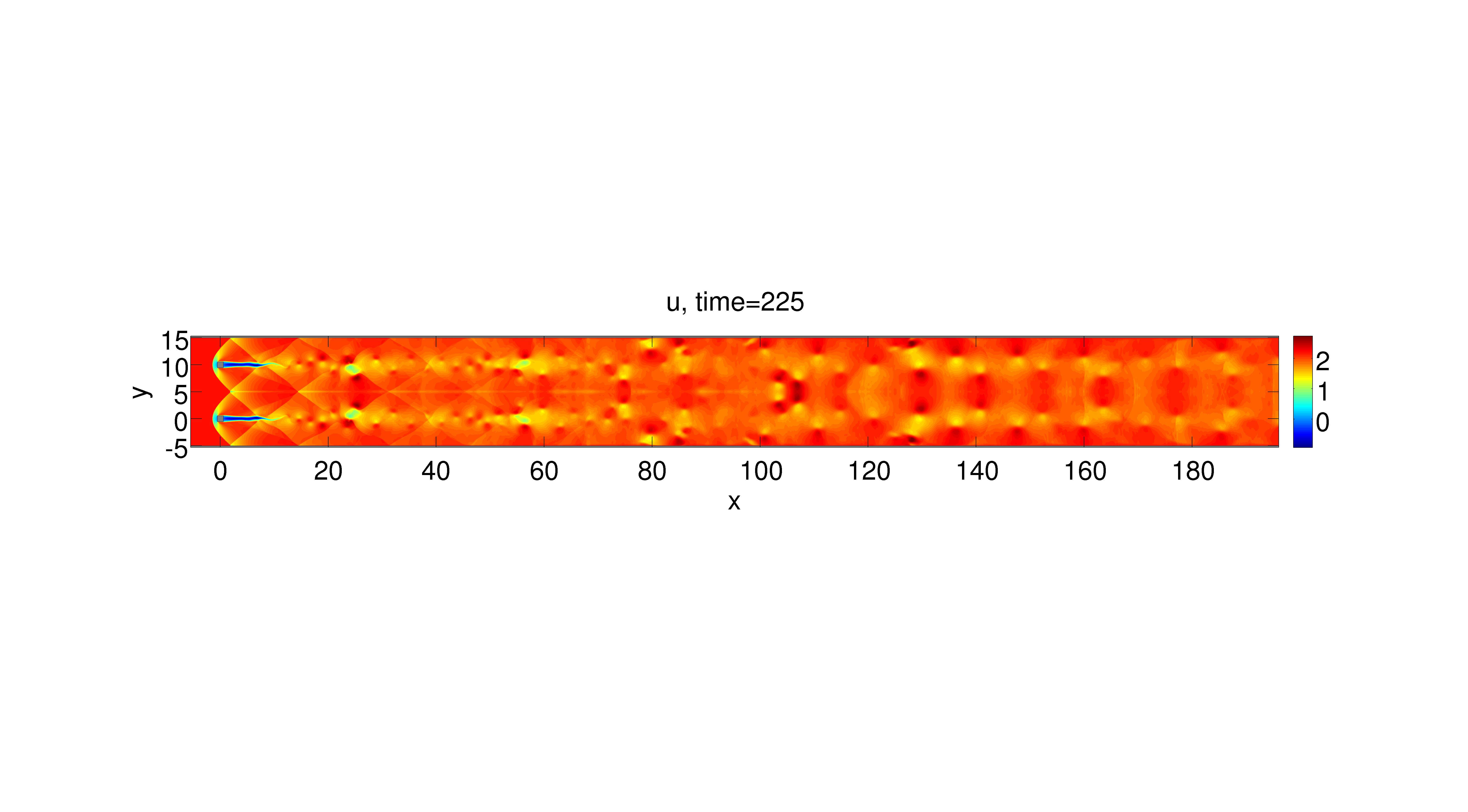}
 \includegraphics[trim=600 1320 580 1400,clip=true,width=1.0\textwidth]{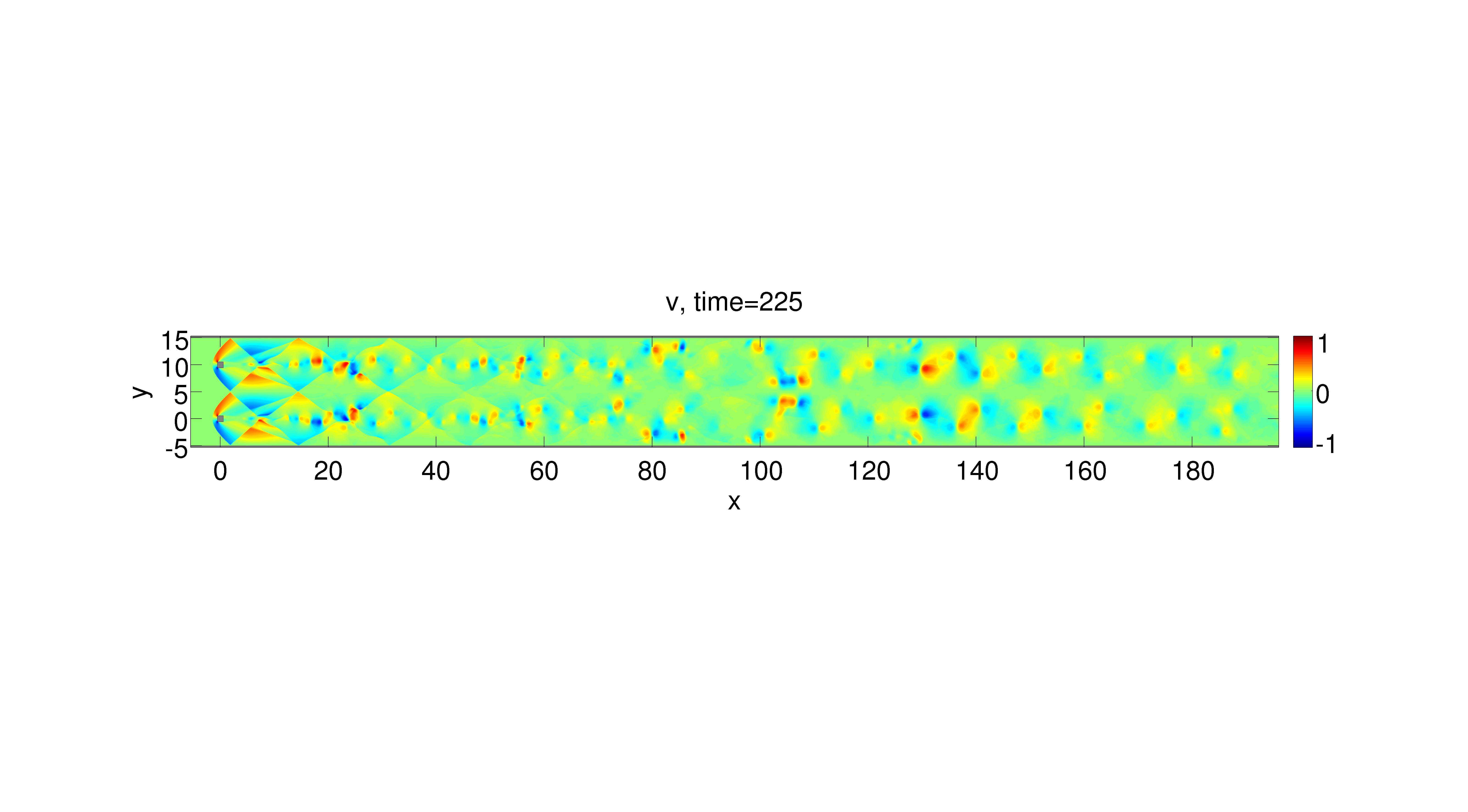}
\caption{The pressure, temperature, horizontal velocity and vertical velocity fields from top to bottom, respectively, calculated by GPU version of \mbox{SIMPLE-TS}.}
\label{uT_fields}
\end{figure}
Here was investigated influence of the number of rows per subdomain over performance because this is an important idea of proposed approach. To this aim, the speedup was obtained on four cases with different channel heights ($H_{ch} = 10, 20, 100$ and $200$) that corresponds to the number of square particles along the x-axis: 1, 2, 10 and 20, respectively. Fig. \ref{uT_fields} shows fields of pressure, temperature, horizontal and vertical velocities for test case with $H_{ch} = 20$ and two square particles. The test cases meshes were 4032x200, 4032x400, 4032x2000 and 4032x4000 points that corresponds to $H_{ch} = 10, 20, 100$ and $200$, respectively. Each work group has 256 work-items (threads) that calculate 252 nodes along the x-axis and four work-items that calculate halo region of the subdomain. Therefore, meshes per subdomain that correspond to test cases are 256x100, 256x200, 256x1000 and 256x2000. $N_y^{subdomain}$ is number of control volumes along the y-axis per subdomain that for calculated test cases $N_y^{subdomain}$ is equal to $100, 200, 1000$ and $2000$. The speedup results were normalized according execution time of serial code on CPU for the corresponding case. Fig. \ref{Speedup_explicit_TVD_GPU_to_CPU}, Fig. \ref{Speedup_explicit_upwind_GPU_to_CPU}, Fig. \ref{Speedup_implicit_TVD_GPU_to_CPU} and Fig. \ref{Speedup_implicit_upwind_GPU_to_CPU} shows the influence of the number of rows of subdomain over the performance. AMD Radeon R9 280X increase speedup with increase of $N_y^{subdomain}$. The obtained performances for $N_y^{subdomain}=1000$ and $2000$ are close; therefore we consider that $N_y^{subdomain}=1000$ is sufficient to reach maximum performance on AMD Radeon R9 280X. The performance of NVIDIA Tesla M2090 does not depend on $N_y^{subdomain}$ and we consider that $N_y^{subdomain}=100$ is sufficient to reach maximum performance of this device.\\\indent
The performance tests show that AMD Radeon R9 280X is significantly faster than NVIDIA Tesla M2090, CPU Intel Core i7-920, and CPU Intel Core i5-4690. The GPU code executed on AMD Radeon R9 280X is faster compared to CPU serial code executed on Intel Core i7-920 from 150x to 184x times. Also, it is faster compared to CPU Intel Core i5-4690 from 81x to 102x times, see Fig. \ref{Speedup_explicit_TVD_GPU_to_CPU}, Fig. \ref{Speedup_explicit_upwind_GPU_to_CPU}, Fig. \ref{Speedup_implicit_TVD_GPU_to_CPU} and Fig. \ref{Speedup_implicit_upwind_GPU_to_CPU}. NVIDIA Tesla M2090 speedup the GPU code compared to serial CPU code executed on Intel Core i7-920 from 9x to 20x times. Also, it is faster compared to CPU Intel Core i5-4690 from 5x to 11x times, see Fig. \ref{Speedup_explicit_TVD_GPU_to_CPU}, Fig. \ref{Speedup_explicit_upwind_GPU_to_CPU}, Fig. \ref{Speedup_implicit_TVD_GPU_to_CPU} and Fig. \ref{Speedup_implicit_upwind_GPU_to_CPU}. The performance of CPU core of Intel Core i5-4690 overcome approximately two times previous generation CPU core of Intel Core i7-920. AMD Radeon R9 280X is the fastest device: it is approximately one order faster than NVIDIA Tesla M2090 and approximately two orders faster than Intel's CPUs Core i5-4690 and Core i7-920. NVIDIA Tesla M2090 is approximately an order faster than Intel's CPUs Core i5-4690 and Core i7-920.\\\indent
\FloatBarrier
\section{Conclusions}
GPU algorithm SIMPLE-TS calculates Navier-Stokes-Fourier system of partial differential equations describing unsteady, viscous, compressible, heat-conductive gas flows with double precision accuracy. A test case was unsteady flow past a square particles in a microchannel at speed $M = 2.43$ and rarefaction $Kn=0.001$.\\\indent
The appropriate use of device memories is important when porting CPU code to GPU. As the private memory is the fastest GPU memory, we use it to keep calculated variables. In local memory were stored temporary calculated arrays that reduce the use of global memory and increase code performance. The equations were put together using macros. As a result, preprocessor composes big expressions that increase Instruction Level Parallelism (ILP). Almost all optimization of the code was left to the compiler. The compile eliminates common subexpression (CSE), organize data copy from global to private device memory and data reuse. The automatic optimization by a compiler improve code maintenance: simplifies code writing and further modifications. The proposed approach demonstrates excellent performance on AMD gaming GPU AMD Radeon R9 280X that overcome one order server NVIDIA Tesla M2090 and two orders serial C++ code run on CPU Intel Core i5-4690 and Intel Core i7-920. After all GPU code obtains excellent speedup on AMD GPU and looks more suitable to AMD GPU architecture than NVIDIA GPU architecture.\\\indent
An important performance tests would be on AMD FirePro W9100 and V100 GPU Accelerator (Mezzanine). AMD FirePro W9100 double precision compute performance is 2.62 TFLOPS that is 2.62 times more than used here AMD Radeon R9 280X. V100 GPU Accelerator (Mezzanine) double precision compute performance is 7.45 TFLOPS that is 11.2 times more than used here NVIDIA Tesla M2090 and could contain important hardware changes.\\\indent
An important demonstration would be the calculation of 3D fluid flow in complex geometry that would establish the performance in a realistic engineering settings.\\\indent

\ \\\indent
\textbf{Acknowledgments}\\\indent
\ \\\indent
We would like to acknowledge the financial support provided by the Bulgarian NSF under Grant DN-02/7-2016. This research has received funding from the European Union Horizon 2020 research and innovation programme under the Marie Sklodowska-Curie MIGRATE grant agreement No. 643095. The calculations on GPU AMD Radeon R9 280X and CPU Intel Core i5-4690 in this work used the cluster of Institute of Mechanics - BAS. The calculations on GPU NVIDIA Tesla M2090 in this work used the EGI Infrastructure and were co-funded by the EGI-Engage project (Horizon 2020) under Grant number 654142.\\\indent

\newpage
\begin{appendices}
\chapter{\textbf{Appendix A.}}\\\indent
\ \\\indent
In this appendix is presented pseudo code of loop along the y-axis. That is the main part of loop 2 of GPU algorithm SIMPLE-TS (see Fig. \ref{SIMPLE-TS_for_CPU_and_GPU_explicit} and Fig. \ref{SIMPLE-TS_for_CPU_and_GPU_implicit}). For brevity, the calculations are presented as function of constant and calculated variables in loop along the y-axis that correspond to the equations (\ref{pl29_1}) - (\ref{pl29_8}) and Fig. \ref{GPU_dependent_variables_calculation} notations.
\begin{verbatim}
//Loop along the y-axis
for(j_global = j_begin; j_global < j_end; j_global++)
{
  //Calculate dencity in middle points rho_u, numerical equation (12)
  //Set row index for the next expression for calculation
  j = j_global + 2;
  if(control_volume_in_fluid(i,j))
  {
    //The calculated control volume is not next to body surface, where
    //boundary conditions have to be applied. Therefore, calculate numerical
    //equation without implemented boundary conditions.
    //Write the result in local memory.
    rho_u(i,j) = f(constant variables in loop 2);
  }
  else if(control_volume_on_the_wall(i,j))
  {
    //The calculated control volume is next to the solid surface, therefore,
    //in numerical equations are implemented boundary conditions.
    //All checks using in implementation of boundary conditions in common
    //case increase number of floating point operations approximately twice.
    //Write the result in local memory.
    rho_u(i,j) = f(constant variables in loop 2);
  }

  //Calculate dencity in middle points rho_v, numerical equation (13)
  //Set row index for the next expression for calculation
  j = j_global + 3;
  if(control_volume_in_fluid(i,j))
  {
    //The calculated control volume is not next to body surface, where
    //boundary conditions have to be applied. Therefore, calculate numerical
    //equation without implemented boundary conditions.
    //Write the result in local memory.
    rho_v(i,j) = f(constant variables in loop 2);
  }
  else if(control_volume_on_the_wall(i,j))
  {
    //The calculated control volume is next to the solid surface, therefore,
    //in numerical equations are implemented boundary conditions.
    //All checks using in implementation of boundary conditions in common
    //case increase number of floating point operations approximately twice.
    //approximately twice.
    //Write the result in local memory.
    rho_v(i,j) = f(constant variables in loop 2);
  }
  //Wait to finish calculations befor continue.
  barrier(CLK_LOCAL_MEM_FENCE);

  //Calculate temperature, numerical equation (27)
  //Set row index for the next expression for calculation
  j = j_global;
  if(control_volume_in_fluid(i,j))
  {
    //The calculated control volume is not next to body surface, where
    //boundary conditions have to be applied. Therefore, calculate numerical
    //equation without implemented boundary conditions.
    //Write result in private and global memory.
    T(i,j) = f(constant variables in loop 2);
  }
  else if(control_volume_on_the_wall(i,j))
  {
    //The calculated control volume is next to the solid surface, therefore,
    //in numerical equations are implemented boundary conditions.
    //All checks using in implementation of boundary conditions in common
    //case increase number of floating point operations approximately twice.
    //approximately twice.
    //Write result in private and global memory.
    T(i,j) = f(constant variables in loop 2);
  }

  //Calculate horizontal pseudo velocity, numerical equation (18) and du
  //Set row index for the next expression for calculation
  j = j_global;
  if(control_volume_in_fluid(i,j))
  {
    //The calculated control volume is not next to body surface, where
    //boundary conditions have to be applied. Therefore, calculate numerical
    //equation without implemented boundary conditions.
    //Write the result in local memory.
    u_pseudo(i,j) = f(constant variables in loop 2);
    du(i,j) = f(constant variables in loop 2);
  }
  else if(control_volume_on_the_wall(i,j))
  {
    //The calculated control volume is next to the solid surface, therefore,
    //in numerical equations are implemented boundary conditions.
    //All checks using in implementation of boundary conditions in common
    //case increase number of floating point operations approximately twice.
    //approximately twice.
    //Write the result in local memory.
    u_pseudo(i,j) = f(constant variables in loop 2);
    du(i,j) = f(constant variables in loop 2);
  }

  //Calculate vertical pseudo velocity, numerical equation (19) and dv
  //Set row index for the next expression for calculation
  j = j_global + 1;
  if(control_volume_in_fluid(i,j))
  {
    //The calculated control volume is not next to body surface, where
    //boundary conditions have to be applied. Therefore, calculate numerical
    //equation without implemented boundary conditions.
    //Write the result in local memory.
    v_pseudo(i,j) = f(constant variables in loop 2);
    dv(i,j) = f(constant variables in loop 2);
  }
  else if(control_volume_on_the_wall(i,j))
  {
    //The calculated control volume is next to the solid surface, therefore,
    //in numerical equations are implemented boundary conditions.
    //All checks using in implementation of boundary conditions in common
    //case increase number of floating point operations approximately twice.
    //approximately twice.
    //Write the result in local memory.
    v_pseudo(i,j) = f(constant variables in loop 2);
    dv(i,j) = f(constant variables in loop 2);
  }
  //Wait to finish calculations befor continue.
  barrier(CLK_LOCAL_MEM_FENCE);


  //Calculate pressure, numerical equation (25)
  //Set row index for the next expression for calculation
  j = j_global;
  if(control_volume_to_calculate == true)
  {
    //Write result in local and in global memory.
    p(i,j) = f(u_pseudo(i,j), du(i,j), u_pseudo(i+1,j), du(i+1,j),\
               v_pseudo(i,j), dv(i,j), v_pseudo(i,j+1), dv(i,j+1),\
               T(i,j), constant variables in loop 2);
  }
  //Wait to finish calculations befor continue.
  barrier(CLK_LOCAL_MEM_FENCE);

  //Calculate horizontal velocity, numerical equation (16)
  //Set row index for the next expression for calculation
  j = j_global;
  if(control_volume_to_calculate == true)
  {
    //Write result in global memory.
    u(i,j) = f(u_pseudo(i,j), du(i,j), p(i-1,j), p(i,j),\
               constant variables in loop 2);
  }

  //Calculate vertical velocity, numerical equation (17)
  //Set row index for the next expression for calculation
  j = j_global;
  if(control_volume_to_calculate == true)
  {
    //Write result in global memory.
    v(i,j) = f(v_pseudo(i,j), dv(i,j), p(i,j-1), p(i,j),\
               constant variables in loop 2);
  }
  //Wait to finish calculations befor continue.
  barrier(CLK_LOCAL_MEM_FENCE);
}
\end{verbatim}

\end{appendices}

\newpage
\textbf{References}
\bibliographystyle{elsarticle-num} 
\bibliography{bibliography_list_1}

\end{document}